%% file: paper.tex
\documentclass[sigconf,nonacm]{acmart}
\setcopyright{none}

\usepackage{chngcntr}
\usepackage{graphicx}
\usepackage{textcomp}
\usepackage{fancyhdr}
\usepackage{todonotes}
\usepackage{float}
\newcommand{\ignore}[1]{}
\usepackage{xspace}
\usepackage{multicol}
\usepackage{caption}
\usepackage{chngcntr}
\usepackage{wrapfig}
\usepackage{subfig}
\usepackage{pifont}
\usepackage{balance}
\usepackage{enumitem}
\usepackage{xcolor}
\usepackage{comment}

\usepackage{csquotes}
\ignore{
\def\csq@setfrcodes{%
    \ifnum\sfcode`\A=\@m
    \else
        \csq@setazcodes
    \fi
    \sfcode`\,=1003
    \sfcode`\;=1004
    \sfcode`\:=1005
    \sfcode`\.=1006
    \sfcode`\!=1007
    \sfcode`\?=1008
}
}
\makeatletter
\DeclareRobustCommand{\change}{%
  \@bsphack
  \leavevmode
  \color{red}%
  \@esphack
}
\DeclareRobustCommand{\stopchange}{%
  \@bsphack
  \normalcolor
  \@esphack
}
\makeatother

\newcommand{\name}{\mbox{{ReplayCache}}\xspace}

\newcommand{\ie}{\textit{i.e.,}\xspace}
\newcommand{\eg}{\textit{e.g.,}\xspace}

\usepackage{todonotes}
\newcommand{\DY}[1]{\todo[color=yellow,inline]{DY:#1}}

\newcommand{\CM}[1]{\todo[color=green,inline]{CM:#1}}

\newcommand{\choiadd}[1]{#1}
\usepackage{soul}
\newcommand{\TODO}[1]{\hl{\mbox{#1}}}
\newcommand{\TODOcite}{\hl{[CITE]}\xspace}
\newcommand{\revision}[1]{#1}

\ignore{
\newcommand\whitecircle[1]{%
\tikz[baseline=(X.base)] 
   \node (X) [
   draw, 
   shape=circle, 
   inner sep=0, 
   fill=white, 
   text=black, 
   font=\sffamily, 
   minimum size=.5em] 
   {\footnotesize #1};}

}

\newcommand*\whitecircle[1]{%
 \protect\begin{tikzpicture}[baseline=(C.base)]                                          
 \protect\node[draw,circle,fill=white,inner sep=0.2pt](C) {\textcolor{black}{#1}};       
 \protect\end{tikzpicture}} 
 
\usepackage{cleveref}
\crefformat{section}{\S#2#1#3}
\crefname{figure}{Figure}{Figures}
\crefname{table}{Table}{Tables}
\crefname{equation}{Equation}{Equations}

\makeatletter
\newcommand*{\rom}[1]{\expandafter\@slowromancap\romannumeral #1@}
\makeatother

\AtBeginDocument{%
  \providecommand\BibTeX{{%
        \normalfont B\kern-0.5em{\scshape i\kern-0.25em b}\kern-0.8em\TeX}}}
\title{Enabling Volatile Caches for Energy Harvesting Systems}
\author{Jianping Zeng}
  \affiliation{
  \institution{Purdue University}
  \country{USA}
}
\email{zeng207@purdue.edu}
\author{Jongouk Choi}
  \affiliation{
  \institution{Purdue University}
  \country{USA}
}
\email{choi658@purdue.edu}
\author{Xinwei Fu}
  \affiliation{
  \institution{Virginia Tech}
  \country{USA}
}
\email{fuxinwei@vt.edu}
\author{Ajay P. Shreepathi}
  \affiliation{
  \institution{Stony Brook University}
  \country{USA}
}
\email{apaddayurush@cs.stonybrook.edu}
\author{Dongyoon Lee}
  \affiliation{
  \institution{Stony Brook University}
  \country{USA}
}
\email{dongyoon@cs.stonybrook.edu}
\author{Changwoo Min}
  \affiliation{
  \institution{Virginia Tech}
  \country{USA}
}
\email{changwoo@vt.edu}
\author{Changhee Jung}
  \affiliation{
  \institution{Purdue University}
  \country{USA}
}
\email{chjung@purdue.edu}

\begin{document}

\begin{abstract}
\input{abstract}
\end{abstract}

\maketitle

\input{introduction}
\input{background}

\input{overview}
\input{region}

\input{recovery}
\input{evaluation}

\input{related-works}

\input{conclusion}
\begin{acks}
The original article was presented in MICRO 2021 \cite{zeng2021replaycache}.
We thank anonymous reviewers for their comments.
At Purdue, this work was supported by NSF grants 1750503
(CAREER) and 1814430. At Stony Brook, this work was 
supported by NSF grant 2029720. At Virginia Tech, this work was
supported by Institute for Information \& communications Technology
Promotion (IITP) grant funded by the Korea government (MSIT)
(No. 2014-3-00035).
\end{acks}

\balance
\bibliographystyle{ACM-Reference-Format}
\bibliography{references}

\end{document}

%% file: abstract.tex
%
Energy harvesting systems have shown their unique benefit of ultra-long
operation time without maintenance and are expected to be more prevalent in the
era of Internet of Things. However, due to the batteryless nature, they suffer
unpredictable frequent power outages. They thus require a lightweight mechanism
for crash consistency since saving/restoring checkpoints across the outages can
limit forward progress by consuming hard-won energy. 
For the reason, energy harvesting systems have been designed with
a non-volatile memory (NVM) only. The use of a volatile data cache has been
assumed to be not viable or at least challenging due to the difficulty to
ensure cacheline persistence.

In this paper, we propose \name, a software-only crash consistency
scheme that enables commodity energy harvesting systems to exploit a volatile data cache.
\name does not have to ensure the persistence of dirty cachelines or record
their logs at run time. Instead, \name recovery runtime \emph{re-executes the
potentially unpersisted stores} in the wake of power failure to restore the
consistent NVM state, from which interrupted program can safely resume. To
support store replay during recovery, \name partitions program into a series of
regions in a way that store operand 
registers remain intact within each region, and checkpoints all registers just
before power failure using the crash consistency mechanism of the commodity
systems.
For performance, \name enables \emph{region-level} persistence that allows the
stores in a region to be asynchronously persisted until the region ends,
exploiting ILP.
The evaluation with 23 benchmark applications show that compared to the
baseline with no caches, \name can achieve about 10.72x and
8.5x-8.9x speedup (on geometric mean) for the scenarios without and
with power outages, respectively.

%% file: introduction.tex
\section{Introduction}
\label{sec:intro}


Energy harvesting systems~\cite{priya2009energy} have been deployed in a wide range of application
domains, such as Internet of Things (IoT) devices~\cite{kamalinejad2015wireless,yau2018energy,
bito2017novel,gorlatova2014movers}, wearables~\cite{chong2019energy,magno2018micro,
magno2017wearable,cheng2015energy,leonov2011energy}, stream and river surveillance
\cite{kamenar2016harvesting,sun2018energy}, health and wellness monitors~\cite{cahill2014structural,
galchev2010vibration,park2008energy,cao2017survey}, etc. Energy harvesting systems are
well-suited to these domains with the superb property of ultra-long operation time without maintenance
by collecting energy from variant ambient sources such as solar, thermal, piezoelectric, and radio-frequency radiation.

However, due to the batteryless nature, energy harvesting systems suffer unpredictable frequent power failure and thus require some form of crash consistency which must be lightweight; otherwise checkpointing/restoring consistent program states across the failure can limit forward progress by consuming hard-won energy. 
Thus, existing systems~\cite{quickrecall,ma2015architecture,balsamo2014hibernus,hicks2017clank,su2016ferroelectric,choi2019cospec,choi2019achieving} have been designed with byte-addressable non-volatile memory (NVM), where data are immediately persisted and thus recoverable at the cost of long latency.
While volatile write-back caches can hide the store latency and improve performance with a load hit exploiting data locality, 
they have been assumed to be not viable or at least challenging in energy harvesting systems.

The crux of the problem is that volatile write-back cache states are not preserved across a power outage.
This may lead to an inconsistent NVM state, and therefore the power-interrupted program may fail to resume correctly.
That is why existing energy harvesting systems do not use volatile data caches; prior work~\cite{ma2015architecture}
uses a read-only NVM-based instruction cache where a crash consistency (without stores) is not an issue.
Unfortunately, it is a challenging problem to ensure correct data cache persistence in a lightweight manner to maintain forward progress. 
For example, software logging causes serious performance degradation (100-300\% slowdown) since each regular store is preceded by the log store, cacheline flush, and store fence~\cite{timestone-asplos20,kolli2016high,wang2018quantify,liu2018ido,volos2011mnemosyne,jeong2020unbounded,jeong2021pmem}.

One possible hardware solution is to use a volatile write-through cache.  It allows
energy harvesting systems to benefit from load hits and to
ensure crash consistency by enforcing that the completion of a store instruction
guarantees the persistence of the data in NVM.  However, write-through cache
comes with a performance penalty on each store as conventional cache-free
energy harvesting processors. Since they use a simple in-order core without any form of speculation, they
cannot hide the data persistence latency.

Alternatively, one can design a persistent write-back data cache, e.g., 
non-volatile cache (NVCache)~\cite{xu2009design,park2012future,jokar2015sequoia,wang20132,
mittal2014writesmoothing,mittal2014lastingnvcache,agarwal2019improving} and non-volatile SRAM cache (NVSRAMCache)~\cite{chiu2010low,li2017design,singh2019design,herdt1992analysis,liu2020low,
sheu2013reram,majumdar2016hybrid}.
However, both cache designs have their own problems. Due to the NVM-based design, NVCaches incur high latency and power consumption for each access.
NVSRAMCaches embed NVM to backup an SRAM-based cache, and checkpoint/restore the entire SRAM to/from the NVM backup across power failure, leading to consume high energy.
While NVSRAMCaches may be as fast as a volatile SRAM cache without power failure,
it is hard to maintain the performance with frequent failure---i.e., the norm of energy harvesting---unless they use a lower-power yet fast non-volatile technology which has not been commercialized yet.


\begin{figure*}[ht!]
  \includegraphics[width=\linewidth]{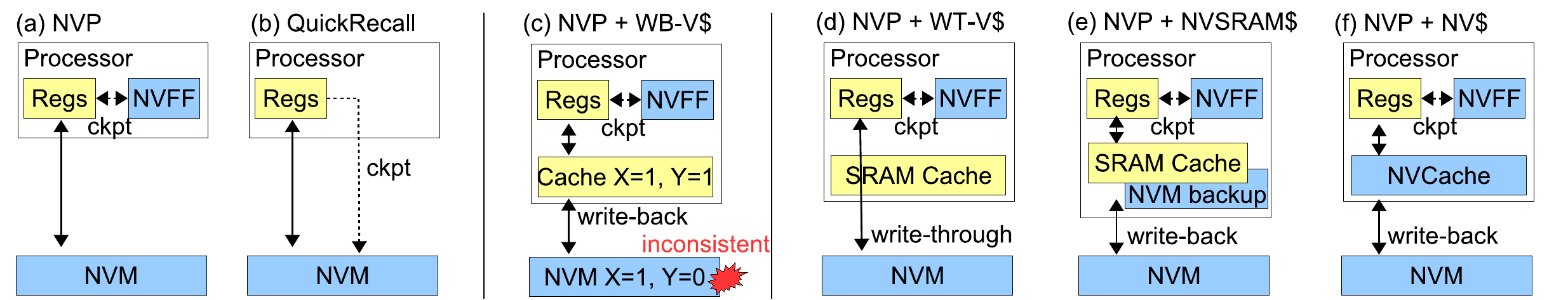}
  \vspace{-15px}
  \caption{The architectures of existing energy harvesting systems 
  \ignore{: (a) NVP~\cite{liu2015ambient}, (b) QuickRecall~\cite{quickrecall}. 
  Yellow boxes are volatile. Blue boxes are non-volatile.
  (c) Using NVP as an example, a naive integration with a write-back volatile cache (NVP + WB-VCache) may lead to an inconsistent NVM state after a power outage.
  Possible workarounds include: (d) NVP + write-through volatile cache (WT-VCache), (e) NVP + NVM-backed SRAM cache (NVSRAMCache), and (f) NVP + NVCache. 
  For (c)-(f), QuickRecall may be used instead of NVP. 
  }
  }
  \vspace{-10px}
  \label{fig:cache}
\end{figure*}

With that in mind, we propose \name, a \emph{software-only scheme} that enables
commodity energy harvesting systems to exploit a volatile write-back data cache
for performance, yet ensures lightweight crash consistency of the NVM state for
correctness. \name does not ensure the persistence of dirty cachelines or
record their logs at run time: \ie no write amplification. Instead, \name
\emph{re-executes the potentially unpersisted stores} in the wake of power
failure to restore the consistent NVM state from which interrupted program can
safely resume.

To support the store replay,
\name partitions program into a series of regions so that the operand registers
of store instructions are intact (\ie not overwritten by the other following
instructions) in each region.
We refer to this process \emph{store-register-preserving region formation}.
Then, at run time, \name checkpoints all registers just before power failure to
secure the store operand registers. We note that the just-in-time register
checkpointing is already available in energy harvesting systems: \eg
QuickRecall\cite{quickrecall}, Hibernus\cite{balsamo2014hibernus}, and
NVP\cite{ma2015architecture}.
%
During recovery, these checkpointed registers are used to \emph{re-execute the
stores} along the same program path as the one before a power failure; for the
store replay, a recovery code block is generated for each region, \ie \name
directs program control to the recovery code in the wake of the power failure.
After that, \name can safely resume from the interrupted program point with the
checkpointed registers and the recovered consistent NVM.



Experiments with 23 applications from Mibench~\cite{guthaus2001mibench} and 
Mediabench~\cite{lee1997mediabench} 
benchmarks show that compared to the baseline with no caches, \name can make them 10.72x and 8.5x-8.9x faster (on geometric mean) for the scenarios without and with power outages, respectively.
This paper makes the following contribution:
\begin{itemize}[nolistsep,leftmargin=10pt]
  \item \name is the first to enable volatile caches for commodity energy harvesting systems; its software-only design allows them to use traditional SRAM cache as is with crash consistency guarantee
  \item \name proposes a new resumption scheme that recovers consistent NVM states across power failure by re-executing potentially unpersisted stores before the failure during the recovery, without write amplification.
  \item \name achieves the high performance despite its software-only design; its performance is comparable to an ideal NVSRAMCache for realistic power failure traces.
\end{itemize}

\ignore{ 
\DY{OLD INTRO}

However, those energy sources are highly dependent on the environmental
conditions and thus lead to low power supplement and unreliability of power supplement. 
With the nature of unreliability in the mind, energy harvesting
systems need to periodically save/restore program status to/from the non-volatile
memory across power outage to make program progress. However, such a
checkpointing/restoring mechanism makes the power outage happened more frequently due to
energy-consuming status checkpointing/restoring or even worse let programs
never get finished~\TODOcite.

To work around the above frequent status checkpointing/restoring process, prior
researches proposed a non-volatile processors (NVPs)~\cite{ma2015architecture,
liu2015ambient, hoseinghorban2019coach} scheme which equips a non-volatile
flip-flops (NVFF) for fast checkpointing/restoring~\cite{wang20123us}. When power
failure happens, the energy buffer is used to supply enough power for saving
all architectural status to the external non-volatile memory (NVM)

When the voltage detector observes that input voltage is about to be below than
a certain threshold, it tells NVFF controller to back up architectural status of
the non-volatile memory. When power comes back, voltage detector signals the
NVFF controller to read checkpointed data from NVM so as to recovery the program
status. After that, NVP can continue the program execution from the program point
where power failure happens.

Even with fast register checkpointing/restoring in the non-volatile
processor, however, there exists high performance overhead (\eg, \TODO{XX}\%)
due to long write
latency characteristic of NVM, such as FeRAM, STT-MRAM (MTJ), PCRAM, and ReRAM 
(memristor) in nvSRAM cells~\cite{chiu2012low,miwa2001nv,masui2003design,
ohsawa20121mb,sheu2013reram,lee2015rram,yamamoto2009nonvolatile}, compared
with traditional DRAM based main memory.

With above performance issue in mind, a natural way to hide the long latency of
store is buffering all stores with caches, which is proved to be one of most
efficient microarchitecture designs for better leveraging program localities.
Nevertheless, naively adapting volatile caches to the NVP leads another
critical issue -- data inconsistency -- when power failure happens or program
crashes because stored data in the non-volatile caches is lost quickly when
power is interrupted. Although this crash consistency problem can be solved
by using volatile cache with write through policy; every store must be written
back to the NVM for persistence before processor continue to execute following
instructions. It is obvious to see that above write through policy leads to
high performance and great amount of power consumption as Section \ref{sec:evaluation}
shows.

The alternative approach to above crash consistency problem is using  non-volatile
memory as a L1 cache, which means data being written by store is preserved once
it has been written back to the non-volatile caches (NVCache). But, non-volatile
caches not only hurts performance (due to expensive write operation~\cite{xu2009design,
park2012future}) but also limits the lifetime of NVCache due to limited write
endurance~\cite{jokar2015sequoia} compared to the traditional SRAM. 

Therefore, we propose \name in this paper to address above challenges without
sacrificing power failure recovery across power outage. More importantly, \name
uses existing SRAM-based cache to achieve high performance. To ensure correct
failure recovery, \name first partitions the program into a series of regions
where \name guarantees registers used by store instructions never get overwritten
by following definitions of the same region. When power failure happens at some
program point of a region, \name replays all stores prior to the failure point
from the last region end by reading register values persisted in NVFF.

With the help of region-level failure recovery, on one hand, \name has the chance
to exploit SRAM with write back policy as a L1 cache instead of either non-volatile
cache or volatile cache with write through policy for achieving better performance.
On the other hand, \name has the ability to avoid store-level persistence. In
other words, \name opts for region-level persistence that only the last store of
region should be acknowledged in the region end; thus the persistence process of
stores in the region could be overlapped with region execution to achieve instruction-level
parallelism (ILP) on top of in-order cores as NVP!

Experimental results with 45 applications from Mibench~\cite{guthaus2001mibench},
Mediabench~\cite{lee1997mediabench}, and EMbench~\cite{embench2019} 
benchmark suites proves \name's high performance compared to the volatile cache
with write through policy, almost 3x performance speedup. The following are our
contributions:

} 

%% file: background.tex
\vspace{-5px}
\section{Background and Motivation}\label{sec:background}

This section discusses the architectures of existing energy harvesting systems (\cref{sec:background:ehs}), the potential crash consistency problem of using a volatile write-back data cache as is (\cref{sec:background:crash}) and the limitations of existing cache solutions (\cref{sec:background:cache}).

\begin{figure*}[t!]
	\begin{center}
    \includegraphics[width=\linewidth]{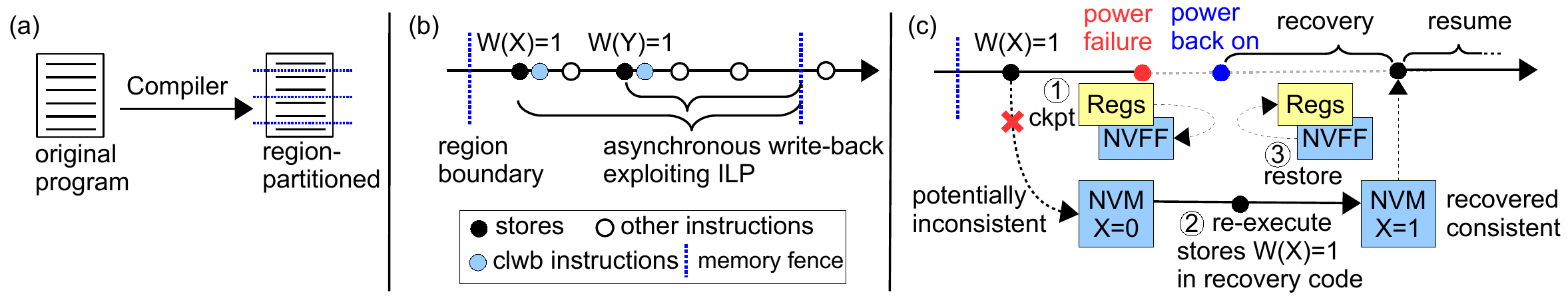}
    \end{center}
    \vspace{-15pt}
    \captionof{figure}{An overview of \name. 
    \ignore{
    (a) \name compiler partitions a program into multiple small regions.
    (b) \name hardware provides region-level persistence, exploiting ILP and ensuring persistence at the region boundary.
    (c) \name \whitecircle{1} checkpoints registers on (before) a power outage. When the power backs on, \name  \whitecircle{2} re-executes stores to recover a consistent NVM state, and \whitecircle{3} restores the register and NVM state to resume.
    }
    }
    \label{fig:overview}
	\vspace{-10px}
\end{figure*}

\vspace{-5px}
\subsection{Architecture of Energy Harvesting Systems}
\label{sec:background:ehs}

Energy harvesting systems derive energy from external sources (\eg solar,
thermal, ambient electromagnetic radiation) and mostly store it in
a tiny capacitor for small IoT devices such as wearables.
Due to the nature of unreliable power supply, energy harvesting systems should
be able to save (checkpoint) the current state upon power failure, and restore
the program state and seamlessly resume the execution when the power comes
backs as if nothing had happened. A power interruption in energy harvesting
systems is
a frequent, normal event, unlike 
in high performance computing context. It is thus crucial to design systems for whole system persistence (WSP)~\cite{narayanan2012whole,kolli2017architecting} so that they efficiently save/restore the program state and make a progress no matter where power failure happens.

The above requirements motivate existing energy harvesting systems to adopt NVM as main memory.
However, the registers in a processor still remain volatile for performance reasons.
Broadly speaking, existing mechanisms to checkpoint/restore registers can be classified into two groups.

\cref{fig:cache}(a) shows the architecture of Non-Volatile Processor (NVP)~\cite{liu2015ambient}, representing the first group that checkpoints and restores registers in place with some additional hardware support~\cite{liu2015ambient,su2016ferroelectric,lee2017reram}.
NVP is equipped with an energy harvester, a voltage monitor, and capacitors (not shown).
When the monitor detects impending power failure, \ie the voltage is about to drop below a certain threshold,
it signals the processor to checkpoint all the registers (so-called just-in-time checkpointing) into their neighboring non-volatile flip-flops (NVFF)~\cite{portal2014overview,onkaraiah2012bipolar,
na2013comparative,li2018lowering}.
When power is secured enough across the failure, the processor restores the register states from the NVFF and resumes the execution from the power-interruption point.
As both register and memory states on the resumption point are guaranteed to be the same as the states before a power failure, there is no 
crash consistency problem.
A downside of NVP is the use of additional hardware NVFF.

\cref{fig:cache}(b) illustrate the architecture of
QuickRecall~\cite{quickrecall}, representing the second group that
checkpoints/restores the registers to/from the NVM. Similar to NVP (and
others), QuickRecall also implements just-in-time (JIT) register checkpointing with a voltage monitor and
a capacitor (not shown). When the monitor detects upcoming power failure, it
triggers an interrupt whose handler checkpoints all the registers into the NVM.
When the power comes back, the recovery runtime reads
the checkpointed states from the NVM in order to restore the registers. As in
NVP, QuickRecall (and others~\cite{balsamo2016hibernus++,balsamo2014hibernus} in this group) has no crash consistency issue.
A drawback of QuickRecall is that it should
secure a lot more energy than NVP to atomically checkpoint all registers in NVM before
impending power failure.

\vspace{-5px}
\subsection{Crash Inconsistency of Write-back Caches}
\label{sec:background:crash}

Adding a cache to energy harvesting systems has a high potential to improve their performance (with load hits) and allow them to make more progress for a given energy harvested. 
However, a naive integration of volatile write-back data cache with existing energy harvesting systems (\eg NVP, QuickRecall) for performance, may lead to a crash consistency problem, as depicted in \cref{fig:cache}(c).

Suppose the NVM has the memory state $X=0$ and $Y=0$ initially. 
And suppose a program has a power outage after executing two stores $W(X)=1$ and $W(Y)=1$.
Before the outage, the cache had the updated state $X=1$ and $Y=1$, but the NVM may not, depending on whether the cache lines holding $X$ and/or $Y$ are evicted or not, which is varying according to cache replacement policy and thus unpredictable.
Since the volatile cache state disappears upon a power loss, \ie any unpersisted dirty cacheline is completely lost,
the system may restart from an inconsistent state (\eg $X=1$ and $Y=0$) failing to resume or producing wrong output later. 

\vspace{-5px}
\subsection{Limitations of Existing Cache Solutions}
\label{sec:background:cache}


There are four possible solutions to address the crash consistency problem.
The first approach is to use a write-through cache.
\cref{fig:cache}(d) illustrates a case in which NVP is configured with a volatile write-through cache (a traditional SRAM-based one).
The write-through policy ensures data consistency as the completion of a store instruction ensures the data persistence to NVM.
However, the downside is a long store latency (as in the case without a cache); more precisely, for a write miss, the critical path is lengthened due to the write-allocation policy.
Since most of the energy harvesting systems are designed with a simple in-order processor, it is impossible hide the store latency.

As shown in \cref{fig:cache}(e), the second approach is to equip the processor with the NVSRAMCache that embeds NVM (\eg ReRAM) to traditional SRAM cache for its backup and restoration~\cite{chiu2012low,miwa2001nv,wei2014design,lee2015rram}.
As with NVP, NVSRAMCache also relies on a voltage monitor for just-in-time checkpointing of the SRAM cache. 
When power is about to be cut, NVSRAMCache triggers a copy from SRAM to NVM for all the cachelines.  Along with their restoration, the entire cache backup makes NVSRAMCache consume high energy across power failure. Moreover, NVSRAMCache significantly postpones the booting time due to the high amount of energy that must be secured for failure-atomic cache checkpointing. 
Although researchers attempt to improve the backup latency~\cite{singh2019design,li2017design}, their NVSRAMCaches are more of a forward-looking technology in an ideal form---since none of current non-volatile materials can provide comparable latency to SRAM~\cite{choi2019cospec}.

The third approach is NVCache~\cite{herdt1992analysis,masui2003design} that leverages a pure non-volatile technology as the cache material; see \cref{fig:cache}(f).
Since NVCache usually uses a slight faster NVM technology for the cache than the non-volatile main memory, the NVCache accesses are a lot slower---consuming more energy---than those of traditional SRAM cache. Thus, NVCache-equipped energy harvesting systems only occasionally outperform cache-free systems when there is very high locality.
In sum, the second and third approaches---Figures~\ref{fig:cache}(e) and (f)---are to make a cache itself persistent surviving power failure, but they suffer from their own problems.

Finally, data loggings are another approach to crash consistency in the
presence of a volatile cache. However, they dramatically increase execution time
(or power consumption if implemented in hardware), prohibiting their use in
energy harvesting systems. For example, iDO~\cite{liu2018ido} and
Mnemosyne~\cite{volos2011mnemosyne} incur 100-300\% slowdown, prohibiting their
use in an energy harvesting system. Furthermore, since they only supports crash consistency for a few transactions or failure-atomic sections, additional overheads should be paid for whole system persistence (WSP)~\cite{narayanan2012whole,kolli2017architecting}. 
Similarly, existing WSP schemes for cache-free harvesting systems such as Alpaca~\cite{Alpaca} and
Ratchet~\cite{van2016intermittent} also cause unacceptable
slowdown (60\% - 500\%).  Since they assume no cache, their overheads would be
even worse for cache-enabled systems due to the additional cacheline flush and
fence overhead.

\ignore{ 
\DY{OLD BACKGROUND}

\subsection{Architectural Design of Non Volatile Processors for Energy Harvesting Systems}
\begin{figure}[htb!]
  \includegraphics[width=1.0\columnwidth]{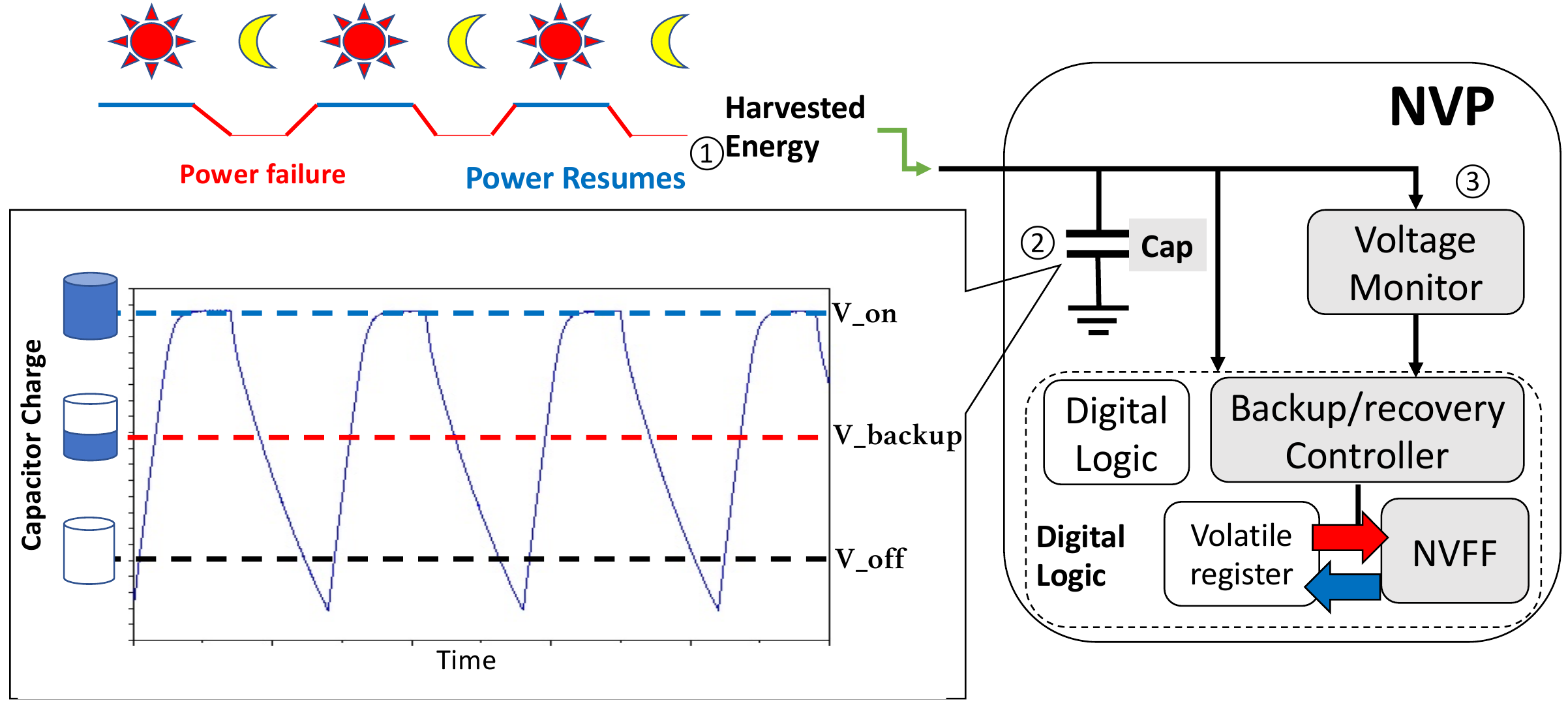}
  \caption{The architecture diagram of energy harvesting non-volatile processors (NVPs).}
  \label{fig:nvp_arch_diagram}
\end{figure}

Due to the nature of unreliable power supply of energy harvesting system,
energy harvesting system needs to frequently checkpoint and roll back program
status to the main memory in case of power failure and power on. However, such
a status saving/restoring process not only causes significant performance overhead
-- frequent status saving/restoring without making program progress -- but also
waste the hard-won energy.

To address above challenge, non-volatile processor (NVP) is proposed to perform
fast register checkpointing and restoring in place with some hardware supports.
As shown in Figure~\ref{fig:nvp_arch_diagram}, an energy harvesting system
(\whitecircle{1}) is equipped to charge the capacitor (\whitecircle{2}), when the voltage
monitor (\whitecircle{3}) detects the voltage is going to drop below the certain threshold,
it signals the processor to take a action of checkpointing all register values
to the non volatile flip-flops (NVFF)~\cite{portal2014overview,onkaraiah2012bipolar,
na2013comparative,li2018lowering} with which volatile register values can be quickly
saved/recalled~\cite{liu2015ambient,su2016ferroelectric,lee2017reram}. When power
is secured enough to complete the register restoration, voltage monitor signals
the processor to restore architectural status from checkpointed register values
in NVFF. In this way, NVP always has the ability to checkpoint and restore the
whole architectural status across power outage without worrying about crash
consistency problem~\cite{liu2015ambient,ma2015architecture,choi2019cospec,wang20123us,lee2017reram}.

\subsection{Limitations of Existing NVP Approaches}
\CM{I suggest to re-org this subsection in this way:
(1) intro to the limitation of NVP w/o cache,
(2) intro to the limitation of NVP w/ non-volatile cache,
(3) challenges in adopting a volatile cache to NVP (in a separate subsection)
}
However, above solution leads to expensive hardware cost, for example, non volatile
flip-flops that should be laid out right next to the volatile registers for fast data
move in between on data saving/restoring process, extra voltage monitor and NVFF
controller, and energy capacitor.
\CM{extra voltage monitor and energy capacitor are common to all energy harvesting systems?}
Even worse, to ensure enough energy for complete
checkpointing, NVP requires more energy consumption, 2$\sim$3X compared to the previous
generations of volatile processors~\cite{wang20123us,liu2015ambient}, which is
caused by higher required input voltage (2.7V$\sim$3V)~\cite{lee2015powering,su2016ferroelectric,
wang2015storage} compared to the prior approaches.

\begin{figure}
  \centering
  \includegraphics[width=0.35\columnwidth,scale=0.1]{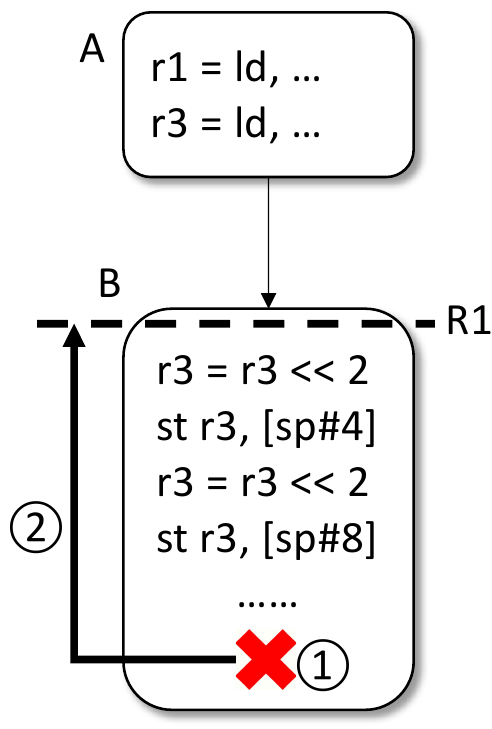}
  \caption{Potential crash consistency issue caused by re-executing an interrupted
           region.}
  \label{fig:potential_crash_consistency}
\end{figure}

NVP is still slower than normal embedded systems with stable power supply even with
fast register values backup/recall provided by NVFF. An intuitive way to boost the
performance of NVP is adapting caches to the non volatile processors as proved over
past few decades. However, it is not always beneficial to adapt caches to the NVP due
to power-hungry nature of cache and potential crash consistency issue. Considering a 
scenario as shown in Figure~\ref{fig:potential_crash_consistency}, after power failure
happened in the bottom of basic block $B$ (\whitecircle{1}), all the data resided in
volatile cache lose.
\CM{Registers will be restored from NVFF?}
To recover from power failure, NVP needs to re-execute (\whitecircle{2})
all stores from region beginning till the failure points. However, such re-execution
process might lead to incorrect program output; register $r3$ value changed when first
store in the line 5 gets re-executed, or even worse program crash when corrupted data
is used to access other memory location.
\CM{Need to add line number to Fig 2}
Hereby, if some registers, \eg $r3$ in Figure
\ref{fig:potential_crash_consistency} used by stores have been overwritten since
the last safe checkpoint; thus, we should not restart the program execution from last
checkpoint. 

One solution to above incorrect program re-execution is leveraging the write through
mechanism of caches to guarantee every store must be persisted in NVM before NVP
continues the execution of following instructions. Therefore, after power comes back,
NVP can simply continue to execute the instruction where power failure happens without
corrupting the program status.

However, to guarantee correct failure recovery protocol, write through policy of volatile
cache significantly degrade the performance as will show in Section~\ref{sec:evaluation}.
With the inspiration of non volatile register files, one might want to use non volatile
caches, \eg NVCache~\cite{xu2009design,park2012future,jokar2015sequoia,wang20132,
mittal2014writesmoothing,mittal2014lastingnvcache,agarwal2019improving},
NVSRAM~\cite{chiu2010low,li2017design,singh2019design,herdt1992analysis,liu2020low,
sheu2013reram,majumdar2016hybrid}, to keep all written data in the cache even after power loss.

Nevertheless, using intrinsically non-volatile memory (NVM), such as ReRAM, PCM, and
STT-RAM, leads to zero standby leakage power, but is constrained in use as an SRAM
due to limited write endurance, access speed. In contrast, NVSRAM---a 3D stacking of
SRAM over non volatile memory---which maintains a non volatile copy bit of each CMOS
bit of SRAM in the underlying NVM when power failure happens. However, the energy
consumed in existing NVSRAM to backup and restore from NVM is high, \eg $\sim$200
$fJ$ as the state-of-the-art work shows~\cite{lee2015rram}. The main reasons are:
(1) high static current for backup and restore process; (2) long backup time to
survive variations; (3) long recovery time for dealing with supply power ramp. As a
result, the high energy consumption nature of NVSRAM prevent it from being used in
energy harvesting systems even though NVSRAM has the ability to address the crash
consistency problem and achieve high performance.

\subsection{Observations and Insights}
\CM{
For non-expert reviewers, we need to explain why we do care performance before explaining our design.
We need to clarify that it is not only simply calculating faster but also making more progress.
}
Due to the correctness problem of using volatile cache with write back policy, performance
problem of NVCache, and power consumption of NVSRAM, it is challenging to utilize those
techniques in the non volatile processors.

The main insight here is we can built up a high performance NVP with a write through volatile
cache if we can guarantee that no store registers has been overwritten from last checkpoint
point till the failure point so that NVP can recover from power failure by selectively
replaying some stores with those registers values checkpointed in NVFF right after power
failure happens.

} 

%% file: overview.tex


\vspace{-5px}
\section{Overview of \name}
\label{sec:overview}


\begin{figure}[t!]
	\begin{center}
    \includegraphics[width=\linewidth]{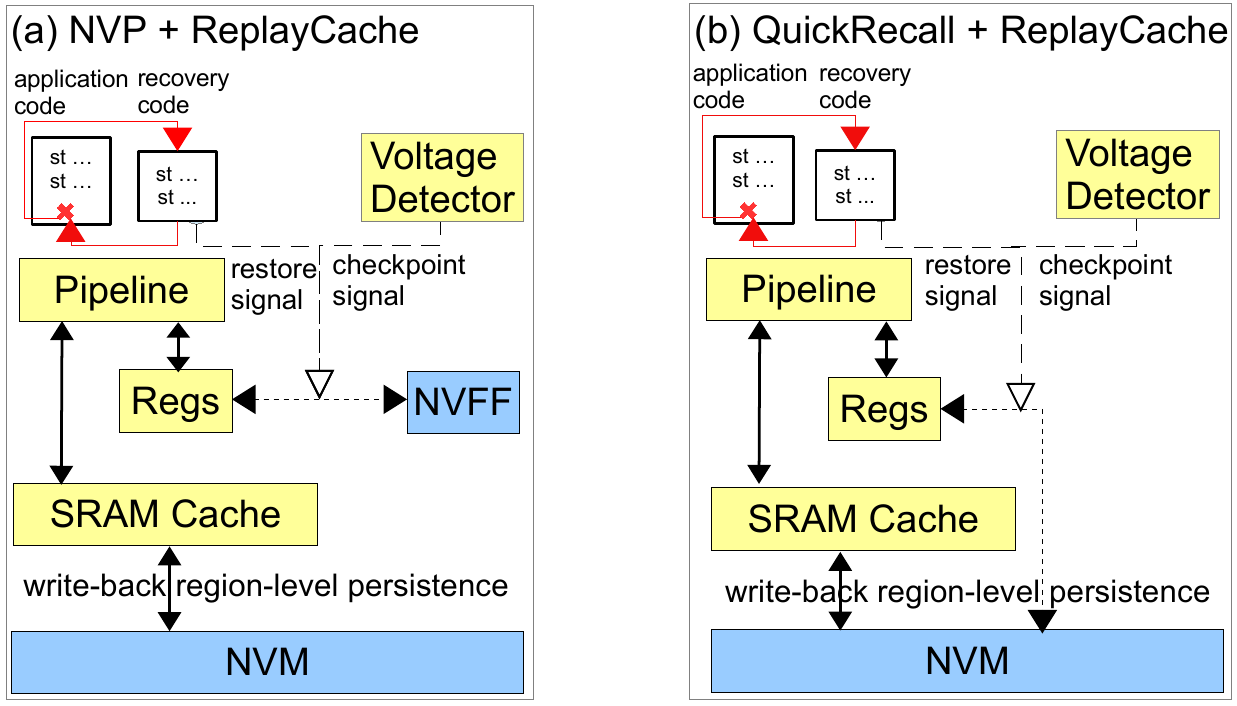}
    \end{center}
    \vspace{-15pt}
    \captionof{figure}{The \name architecture based on a volatile write-back cache. \name can be combined with existing energy harvesting systems
    such as (a) NVP and (b) QuickRecall. Voltage detector is available in every energy harvesting system.
    }
    \label{fig:arch}
\end{figure}

\begin{figure*}[t!]
  \includegraphics[width=\linewidth]{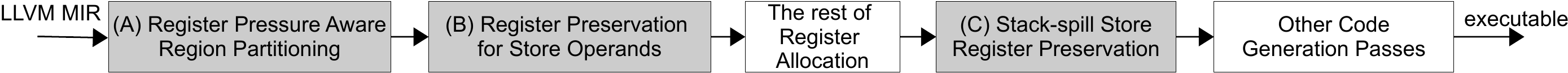}
  \caption{The workflow of \name compiler.
  }
  \label{fig:region_formation}
  \vspace{-15px}
\end{figure*}

The goal of \name is to guarantee crash consistency (\ie an ability to restart from a consistent state) of energy harvesting systems in the presence of a volatile write-back data cache, allowing them to make the most of data locality and to achieve more progress given an energy budget.
\name employs software-only design that provides (A) program region partitioning, (B) region-level persistence, (C) register checkpointing before a power outage, and (D) recovering a consistent NVM state.

\subsection{Program Region Partitioning}

As shown in \cref{fig:overview}(a), \name compiler partitions entire program input to a series of regions. 
Each region ensures that the operand registers (\eg address, value) of a store therein are not overwritten by any other succeeding instructions in that region.

\subsection{Region-level Persistence}\label{sec:region_level_persist}
\name asynchronously writes back the stored value to the NVM, and overlaps the write-back operations with the executions of other following instructions, effectively exploiting instruction-level parallelism (ILP).

Unlike a traditional write-back cache, \name ensures that all the stores in a region are persisted (written back to the NVM) before the region ends; this paper calls this \emph{region-level} persistence guarantee in which the persistence latency of in-region stores can be naturally hidden by ILP; \cref{fig:overview}(b) illustrates the window of potential ILP gain, and the unpersisted state of each store.
This region-level persistence assures that at the moment of a power outage, all the stores in the preceding program regions have already been persisted, and only the stores in the interrupted region could not potentially be unpersisted.

The processor stalls if there exists an outstanding unpersisted store at the end of a region, until it becomes persisted to the NVM.
\name compiler dedicates a single register (\eg $r12$) to be acted as \emph{region register} to track the most recent region boundary information for recovery. That is, the register is updated with a program counter at each region boundary.

\subsection{Register Checkpointing}


Across a power outage, \name saves register states just before the outage and
restores them in the wake of the outage using the voltage monitor based JIT
checkpointing mechanism (\cref{sec:background:ehs}) in commodity energy
harvesting systems. For instance, NVP and QuickRecall can both checkpoint register
states before the power off and to restore them after the power on as discussed
in~\cref{sec:background:ehs}. In \cref{fig:overview}(c), step \whitecircle{1}
illustrates that \name checkpoints the registers when power is about to be cut
off.

\subsection{Power Failure Recovery}

The recovery protocol works as follows. Upon a power outage, the interrupted region's
stores before the outage may or may not be persisted, \eg $W(X)=1$ in
\cref{fig:overview}(c) unpersisted till the outage---while all 
preceding regions' stores are guaranteed to be persisted and thus
consistent (due to the region-level persistence).
In the wake of the outage, \name jumps to the recovery code block of
the interrupted region to replay all the stores left behind the outage.
The recovery code block re-executes such unpersisted stores using the checkpointed
register values in either NVFF (NVP) or NVM (QuickRecall). This is shown as
a step \whitecircle{2} of \cref{fig:overview}(c). Finally, the recovery code
sets off a restoration signal to restore all registers (including PC) from NVFF
or NVM, and then resumes the program from the outage point with the restored
register and the recovered NVM states as in step \whitecircle{3} of \cref{fig:overview}(c).
In this way, \name allows energy harvesting systems to seamlessly leverage
a data cache without amplifying NVM stores. 

\cref{fig:arch} depicts how \name works for existing energy harvesting systems,
\ie NVP and QuickRecall, using the aforementioned recovery protocol. The
takeaway is that \name enables the commodity systems to leverage write-back
volatile data caches as is with help of the region-level persistence and the
recovery code based recovery.  The details of recovery code block generation
is presented in Section~\ref{sec:recovery}.

%% file: region.tex
\vspace{-5px}
\section{\name Compiler}
\label{sec:region}
This section describes how \name compiler realizes the
\emph{store-register-preserving region formation}. The compiler's role is 3-fold: (1) region formation (2) CLWB insertion after each store, and (3) recovery code generation whose discussion is deferred to Section~\ref{sec:recovery}.

For region formation, the compiler partitions program
into a series of small regions so that in each region, no operand registers of a
store instruction are overwritten by the following instructions.
%
%
That way store registers remain intact
from the execution of their region all the way to the power failure recovery
time on which \name replays the same stores in case they were not persisted before the failure.
We refer to this property as \emph{store integrity}.

Figure~\ref{fig:region_formation} shows a high-level workflow of \name compiler
which introduces 3 additional phases (shaded in the figure) to the standard
backend compilation passes. This region
formation is performed in a whole-program manner to cover the entire program
stores, \ie every single program point belongs to one of the regions.

At first glance, forming regions appears to be as simple as
counting the store registers while traversing the control flow
graph (CFG) and placing boundaries before the count exceeds the
number of (physical) registers in the processor (\eg 16 for NVP and QuickRecall).
However, it turns out that two problems below make the region formation challenging.

{\bf Problem 1. Circular Dependence:}
Intuitively, the store-register-preserving region formation can be
realized with two phases: (1) region partitioning that counts
stores to place a region boundary, i.e., store fence instruction, in program and then (2)
register preservation that extends the live interval of store operands
to the end of each region for their exclusive register use.
Thus, the register preservation depends on the region partitioning. 
However, since the partitioning counts the stores to determine where to place
a region boundary, it also depends on the register preservation---forming a circular dependence;
the live interval extension of the register preservation increases the register pressure, \ie the number of necessary registers. 
Due to the register file size limitation, some registers could be spilled (written) to stack through stores. We call them stack-spill stores.

{\bf Problem 2. Stack-Spill Stores:}
In addition to regular stores, \name also needs to ensure the integrity of stack-spill stores for correct failure recovery.
However, it is hard for the region partitioning to figure out in advance what variables are to be spilled to stack.
That is because stack-spill stores are determined in the later register allocation pass assigning physical registers. 
One might try to perform the region partitioning after the register preservation to exactly count the
number of stores. 
However, this is not a viable option since the region partitioning depends on the register preservation in the first place.

{\bf \name Approach to the Problems:} To break the circular dependence between the region partitioning and the register preservation,
\name first considers a function call boundary as initial regions and conducts (A) \emph{register-pressure aware region partitioning} (the first box of Figure~\ref{fig:region_formation}) to fine-cut the initial regions as needed. 
Our register-pressure tracking algorithm allows the region partitioning phase not only to estimate the number of stack-spill stores, breaking the dependence on the register preservation, but also possibly to form a region with no spill in a best-effort manner.
In case register allocation actually generates stack-spill stores in the formed region after the (B) register preservation phase, \name runs a post-processing (C) \emph{stack-store register preservation} phase (the fourth box of the figure) that runs through the register-allocated code to find those stack-spill stores whose registers are overwritten in their region, and places a region boundary before the register updates. The rest of this section details the three phases with referring to them with (A), (B), and (C), respectively.

\begin{figure*}[ht!]
  \centering
  \includegraphics[width=\linewidth]{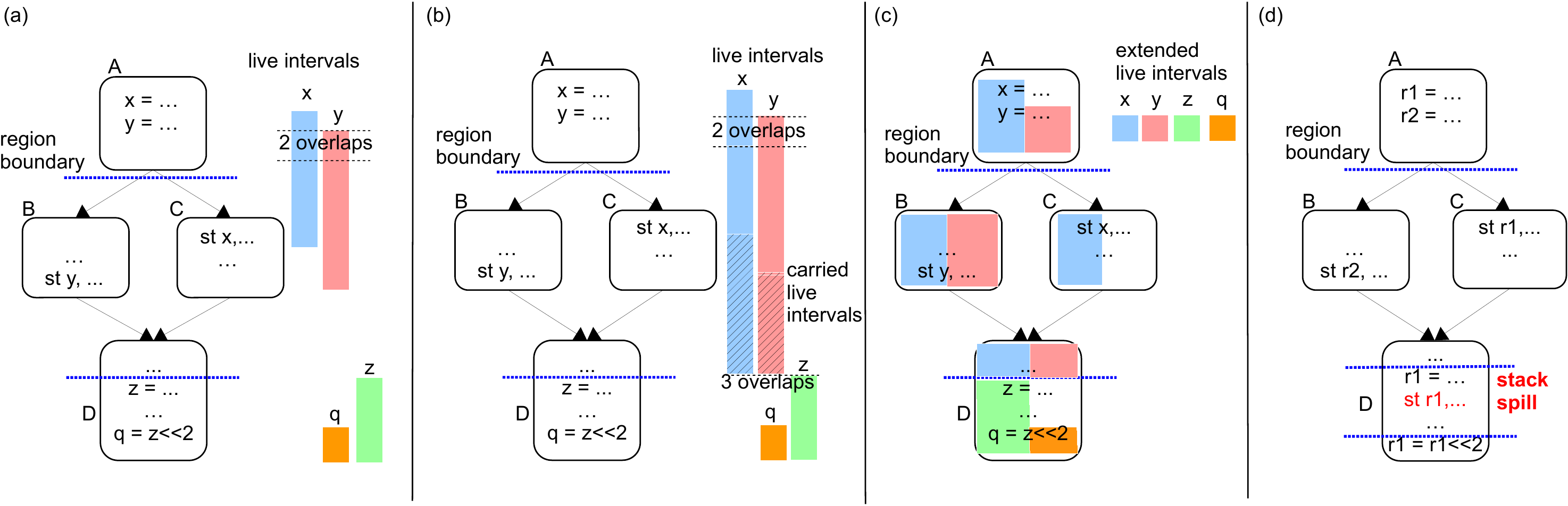}
  \vspace{-18px}
  \caption{An example partitioned program with live intervals;
           (a) shows an initial region boundary at the function beginning
           (basic block $A$) and live intervals of variables $x, y$, and $z$;
           (b) shows the second boundary inserted in basic block $D$ over which
           live intervals of $x, y$ are carried, when the partitioning threshold
           (physical registers) is two; (c) shows extended intervals of all three variables
           towards the second region boundary; and (d) shows the case of
           redefining the register $r1$ of spill store in basic block $D$
           after variables are assigned to the physical register.
           }
  \vspace{-10px}
  \label{fig:live_intervals}
\end{figure*}

\vspace{-5px}
\subsection{Register Pressure Aware Partitioning}
\name initially forms regions at function call boundaries and the end of conditional branches, and then runs the 
register-pressure aware region partitioning algorithm, which aims to achieve two goals.
First, it attempts to maximize the length of a region to provide \name with long potential ILP window for its
region-level persistence; see \cref{fig:overview} (b).
Second, it tries to minimize stack spills generated by the later register allocation phase.

For this purpose, the partitioning algorithm keeps track of the register pressure by
traversing the control flow graph (CFG) of each initial region.
\name counts the number of overlapping live intervals at each
program point visited during the CFG traversal.  In particular, if
store instructions are encountered, \name carries their live
intervals along the way beyond the original live intervals.
This serves as a
proxy for the actual live interval extension of the next (B) register
preservation phase.  When the number
of the overlapping live intervals becomes greater than the number
of physical registers available in the underlying processor, a
stack-spill store might be generated thereafter.  Therefore, a
region boundary, i.e., store fence, is placed at that point. That way \name can maximize
the size of the store-register-preserving region, likely with no spill.

Figure~\ref{fig:live_intervals}(a) shows an example code where there are
variables $x$, $y$, $z$ and their live intervals; $x$ and $y$ are used as store operands, and their live intervals overlap
in basic block $A$ as shown in the left of the figure. Suppose there are only 2 physical
registers. Figure~\ref{fig:live_intervals}(b) demonstrates how the
register-pressure aware region partitioning works for the example code.
Basically, whenever stores are encountered, the algorithm carries the live interval of their operands for the rest of the CFG traversal. 
For example, when the traversal hits the store $y$ at the end point of basic block $B$ in the left control path, the algorithm will start carrying the live interval of $y$ thereafter (illustrated as a hatched box in the figure); the same action is taken with the store $x$ in the right path.
Thus, when the traversal hits the point where $z$'s assignment is found in the
join basic block $D$, the live intervals of both $x$ and $y$ have been carried
to the point. Since $z$'s live interval starts there, the
algorithm places a region boundary at that point, which would otherwise 
end up making the number of overlapping intervals (3 thereafter) bigger than 
the number of physical registers (2).  

\vspace{-5px}
\subsection{Regular Store Register Preservation}
Once regions are formed by the register-pressure aware region partitioning, \name compiler enters register allocation.
Then, this register preservation phase ``preserves'' the variables used for the operands of stores.  The goal is to ensure that no other variables are assigned to
those registers that are supposed to be occupied only by store operands. To achieve this, this phase extends the live interval of
store operand variables from their last use point \emph{to the end of the region} to
which they belong, along the control path.  

For example, as shown in Figure~\ref{fig:live_intervals}(c), the actual live intervals of $x$ stops
at its last use point in basic block $C$, the resulting interval is
extended to the next region boundary placed in the middle of the bottom basic block
$D$; similarly, $y$'s interval is extended to the same following region boundary.  In this way, $x$ and $y$ never share their physical registers---even after their
last use point---with other variables. 
In other words, the next register allocation phase ensures
that neither $x$ nor $y$ is assigned to any physical register used by other variables.
Consequently, \name guarantees the integrity of the regular stores' registers.

\subsection{Stack-Spill Store Register Preservation}
The register
allocation might spill some variable to stack and generate the stack-spill
stores. This actually happens since register allocation performs in a
function level (not a region level) and makes a global decision 
across all the regions in a function---though the (A) register-pressure aware
region partitioning tries to form spill-free regions in a best-effort manner.
Just in case, this stack-spill store register preservation phase searches the
register-allocated code of each region for any update on the spill store registers.
For example, in Figure~\ref{fig:live_intervals}(d), a $r1$ is spilled to the stack
in basic block $D$, \ie the stack-spill store of $r1$ is generated there. However,
in the region, the spill store is followed by the instruction that changes the $r1$,
\ie $r1=r1\ll2$. Thus, the region cannot guarantee the integrity
of $r1$ used by the stack-spill store from that moment. To deal with this problem, 
this phase places an additional region boundary right before
the register updating instruction to separate it from the stack-spill store; the resulting boundary is shown 
near the bottom of basic block $D$ in Figure~\ref{fig:live_intervals}(d). Consequently, \name compiler
guarantees the integrity of all the store registers in all regions.

\noindent{\bf CLWB insertion:} Once register allocation ends, after which
no store is generated, the compiler inserts a CLWB instruction right after
each store in regions. Since CLWB instructions reuse the address operand of
the preceding store, they make no side effect other than the
instruction count increase.

%% file: recovery.tex
\vspace{-5px}
\section{Recovery Protocols}
\label{sec:recovery}


This section describes (A) how \name compiler generates recovery code and (B) the
details of recovery procedure, and (C) finally explains a running example.

\vspace{-5px}
\subsection{Recovery Code Generation}

To recover from power failure, as a software-only design without hardware
support, \name compiler generates a recovery code block for each region, which contains all the
necessary information and code for the recovery of the region. A recovery code block consists of
\emph{Recovery Code}, which is a code to re-execute all stores in the corresponding region, and
two maps---\emph{Recovery Map (RM)} and \emph{Store Counting Map} (CM)---to
locate the corresponding recovery block and the number of stores to be
re-executed for recovery. An RM is a map from a region boundary PC to an
address of region recovery code. A CM is a map from a region boundary PC to
a \emph{Store Counting Table (SC table)}, which is an array of store addresses
and the number of store instructions from the beginning of the region to this store. With these generated
recovery code and maps, \name's recovery protocol figures out where the
recovery code of the interrupted region is and how many stores should be
re-executed in the interrupted region before the failure point.

In particular, to ensure the absence of power failure during the recovery
process, \name compiler leverages the EH model~\cite{san2018eh} to estimate the
worst-case execution energy of the recovery code block. If the energy is
greater than what the underlying capacitor can deliver with it full
capacitance\footnote{Energy harvesting systems do not reboot across power
failure until the capacitor is fully charged, which is the case for commodity
systems such as NVP, WISP, and QuickRecall.}, the compiler splits the
corresponding region into two smaller regions and generate their recovery
code blocks; this process is repeated unless the resulting code blocks are
small enough to complete with the fully charged capacitor.  In this way,
\name guarantees the power-failure-free recovery.  According to
experimental results (\cref{sec:eval}), \name regions are not that long; we
have not encountered any regions that must be split during our evaluation
of total 23 benchmark applications.

\vspace{-5px}
\subsection{Recovery by Re-execution}
\name's region-level persistence guarantees that all the stores in preceding
regions are persisted. However, stores in the interrupted region before the
power outage may or may not be persisted. \name recovery protocol relies on two
properties: First, upon power outage, \name processor checkpoints registers
(including PC) just-in-time by signaling voltage monitor (NVP) or runtime
(QuickRecall). The register checkpoint is thus available in either NVFF (NVP)
or checkpointing storage in NVM (QuickRecall). Next, \name compiler ensures that
registers used for store operands are never overwritten within a region. This
implies that \name can restore memory status from potential corruption by
re-executing the recovery code generated by the compiler.

\begin{figure}[t!] \begin{center}
\includegraphics[width=1.0\linewidth]{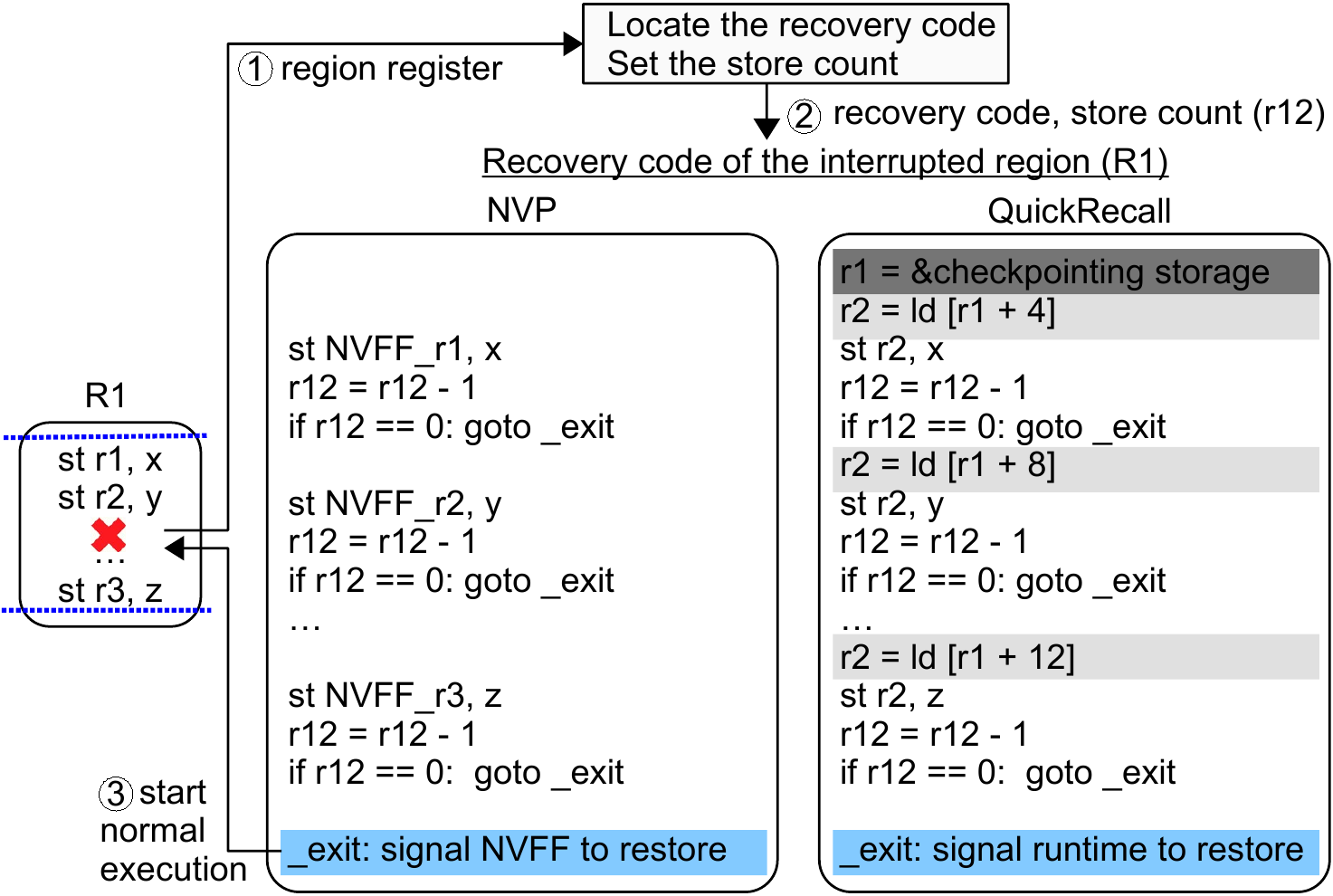} \end{center}
\vspace{-15pt} \captionof{figure}{
    Failure recovery of region $R1$ when an outage happens in the middle
    of basic block $A$. Upon recovery, \name locates a recovery code and counts
    the number of stores needed to be re-executed \whitecircle{1}. Then it
    re-plays all stores in the recovery block by using checkpointed store
    operand registers in NVFF \whitecircle{2}. Finally, it goes back to the failure
    point by restoring registers from NVFF and continues the normal
    execution \whitecircle{3}. 
    } \label{fig:example}
\end{figure}

When the power comes back, \name first finds out the start address of an
interrupted region. It loads the checkpointed \emph{region register} --
a dedicated general-purpose register by compiler as mentioned in
\cref{sec:region_level_persist} -- from NVFF or checkpointing storage, and locates
the recovery code and the SC table of the interrupted region by looking up the RM
and CM, respectively. \name gets the number of store instructions to be
re-executed from the beginning of the region to the failure PC by performing
binary search of the SC table with the region register as a key. Subsequently,
\name runtime jumps to the recovery code of the interrupted region with the
re-executing store count in a register. As illustrated in
Figure~\ref{fig:example}, the recovery code is a series of re-executing the
store instruction, decrementing the store counter, and checking if the counter
is zero. After executing the specified number of store instructions (\ie the
store counter becomes zero), \name runtime signals voltage monitor to restore
register files from either NVFF or checkpointing storage and thus goes back to
the failure point because PC now points to the failure point.

\subsection{A Running Example}

We illustrate a recovery example in Figure~\ref{fig:example}. \name compiler
ensures that registers that are used for store operand ($r1$, $r2$, and $r3$)
are never updated in region $R1$. When entering into $R1$, \name sets the region
register to the beginning of $R1$. When a power outage happens in the region
indicated by a red cross, all registers, including the region register and PC,
are checkpointed. At this point, the stores to memory locations $x$ and $y$ may
or may not be persisted due to the volatile cache.

When the power comes back, \name first loads the region register, which points
to the beginning of the interrupted region. Then it locates the corresponding
recovery code and the number of stores to be re-executed from the RM and SC
table \whitecircle{1}. \name jumps to the recovery code to re-execute the same
number of store instructions in the region before the failure \whitecircle{2}.
In the recovery code examples in Figure~\ref{fig:example}, $r12$ is the number
of stores to be re-executed during recovery. In the recovery code, \name
runtime loads the checkpointed store operand registers (\eg $NVFF\_r1$ in NVM,
and $ld~[r1+4]$ in QuickRecall) and re-executes store instructions. Once \name
runtime re-executes the same number of store instructions -- \ie all store
instructions to the failure PC are re-executed, the store counter
($r12$) becomes zero and the runtime prepares to resume the normal execution
($goto \_exit$ colored in blue). The runtime signals voltage monitor to restore
register files from NVFF and jumps to failure point \whitecircle{3}. The
recovery code are slight different between NVP and QuickRecall. As shown in the
right, QuickRecall loads the checkpointed registers from the storage (colored
in gray).

%% file: evaluation.tex
\section{Evaluation}\label{sec:eval}
\ignore{
We evaluate \name to answer the following:

\begin{itemize}[nolistsep,leftmargin=10pt]

\item How much performance improvement \name can enable for the scenarios with/without power outages? How much better is it compared to the other schemes? (\cref{sec:eval:perf})
\item How much ILP can \name exploit? (\cref{sec:eval:ilp})
\item How much code size does \name increase? (\cref{sec:recovery_code_size})
\revision{\item How many instructions does \name add? (\cref{sec:dynamic_inst_count})}
\item Sensitivity to cache size and NVM technologies? (\cref{sec:eval:sense})
\item How long are the store-register-preserving regions? (\cref{sec:eval:region})
\end{itemize}
}
\subsection{Methodology}
\subsubsection{Compiler}
%
We implemented all \name compiler passes using the
LLVM compiler infrastructure~\cite{lattner2004llvm}. In particular, we implemented our LLVM
passes on MIR (Machine IR) level after instruction selection to precisely
measure the number of live intervals during the region construction. The all compiler
passes consist of about 1700 LOC excluding comments.

\subsubsection{Architecture}
We evaluate \name using a gem5 simulator~\cite{binkert2011gem5} with ARM ISA,
modeling a single core in-order processor with 16 registers, based on the NVPsim~\cite{gu2016nvpsim}; \cref{table:NVM} summarizes our NVM write/read latency based on~\cite{gu2016nvpsim,poremba2012nvmain,liu2019janus,liu2018crash}.
In particular, we only modified L1D cache leaving L1I cache as NVM cache as with the original NVP~\cite{liu2015ambient}.
Note that \name works for any energy harvesting processors that support
just-in-time (JIT) register checkpointing. In addition to NVP, we test \name on
top of QuickRecall whose simulation configuration follows that of NVP other
than the JIT checkpointing/restoration parameters.
\cref{table:comp} shows the detailed simulation parameters of NVP and QuickRecall. 
Since QuickRecall checkpoints registers in NVM, its checkpoint/restore voltage thresholds are higher than those used by NVP.


\subsubsection{Other Cache Designs and the Default Setting}
In addition to \name, we test 3 alternative cache designs: non-volatile cache (NVCache), non-volatile SRAM cache (NVSRAM), and volatile write-through cache (WT-VCache).
All 4 cache designs are assumed to run with NVP unless noted otherwise.
Especially for NVSRAM, we use the same configuration used by
NVPsim~\cite{gu2016nvpsim}, which is based on advanced ReRAM technology. That
is, it writes 3x faster with 5x less energy compared to conventional ReRAM based non-volatile main
memory does. Similarly, it reads 2x faster with 24x less energy compared to the main memory does.
Thus, NVSRAM here serves as the upper bound for performance comparison due to the forward-looking technology used.
As our default setting, we set the size of all the caches to 8KB, and they are all 2-way set-associative. 
For non-volatile main memory, we used Re-RAM by default and set its size as 16MB by leveraging NVMain~\cite{poremba2012nvmain}.
We also perform sensitivity studies with STT-RAM and PCM using the parameters in~\cref{table:NVM}.

\begin{table}[t]
\caption{The timing parameters (ns) of different NVM technologies:
\eg tCK stands for clock period.}
\label{table:NVM}
\centering
\resizebox{\linewidth}{!}{  
\begin{tabular}{|l|l|l|l|l|l|l|l|}
 \hline   NVM & tCK&tBURST&tRCD&tCL&tWTR&tWR&tXAW \\ \hline\hline
ReRAM (default) & 0.94&7.5&18.0&15.0&7.5&150&30 \\ \hline 
STT-RAM         & 1.5&6&35&15&12.5&25&50        \\ \hline
PCM             & 1.88&7.5&48.0&15.0&7.5&300&50 \\ \hline
\end{tabular}
}
\end{table}
 \begin{table}[t!]
	\caption{Simulation configuration.
  }
	\label{table:comp}
	\centering
	\resizebox{\linewidth}{!}{
  \begin{tabular}{|l|l|l|l|}
		\hline
		& \textbf{NVP} (default)  & \textbf{NVP} (NVSRAM)         & \textbf{QuickRecall}    \\ \hline \hline
		Vmax/Vmin\cite{su2016ferroelectric}        & 3.3/2.8   & 3.5/2.8      & 3.5/2.8     \\ \hline
		Ckpt/Restore\cite{su2016ferroelectric}   & 2.9/3.2     & 3.2/3.4      & 3.1/3.3     \\ \hline
		Recovery    & NVFF+Cache   & NVFF+Cache    & VFF+Cache\\ \hline
	\end{tabular}
  }
\end{table}

\subsubsection{Benchmarks and Power Traces}
\begin{figure}[!t]
\centering
\vspace{-10px}
\subfloat[Power Trace 1 (Home)]{\includegraphics[width=0.5\columnwidth, angle=0]{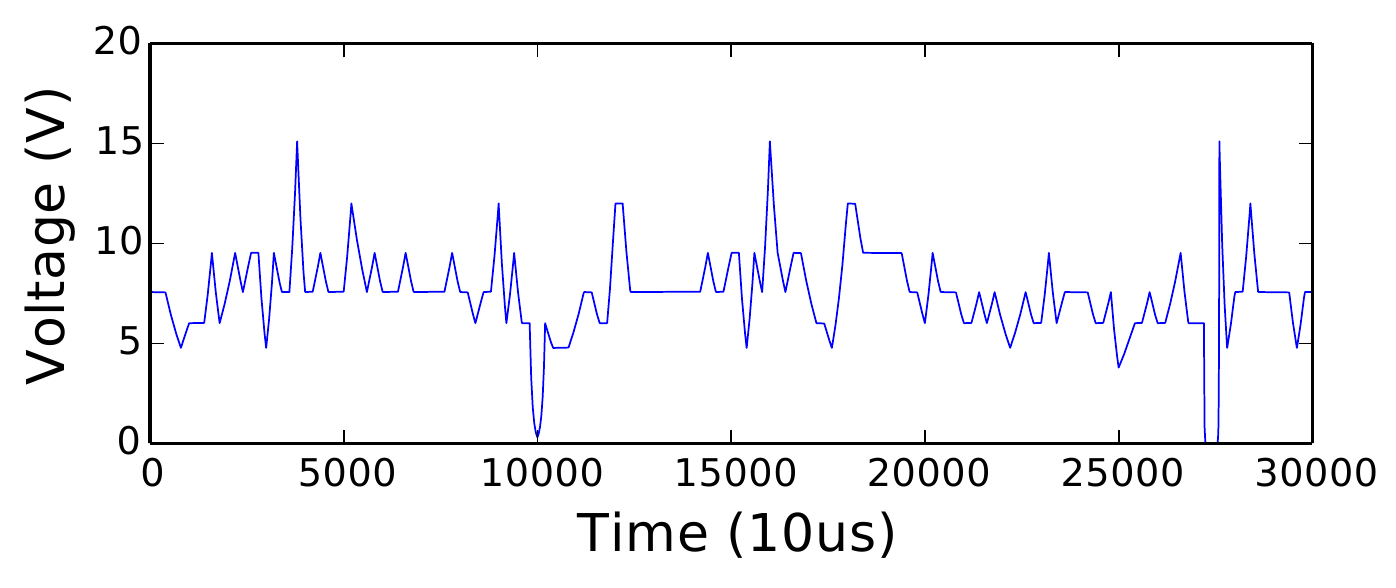}}
\subfloat[Power Trace 2 (Office)]{\includegraphics[width=0.5\columnwidth, angle=0]{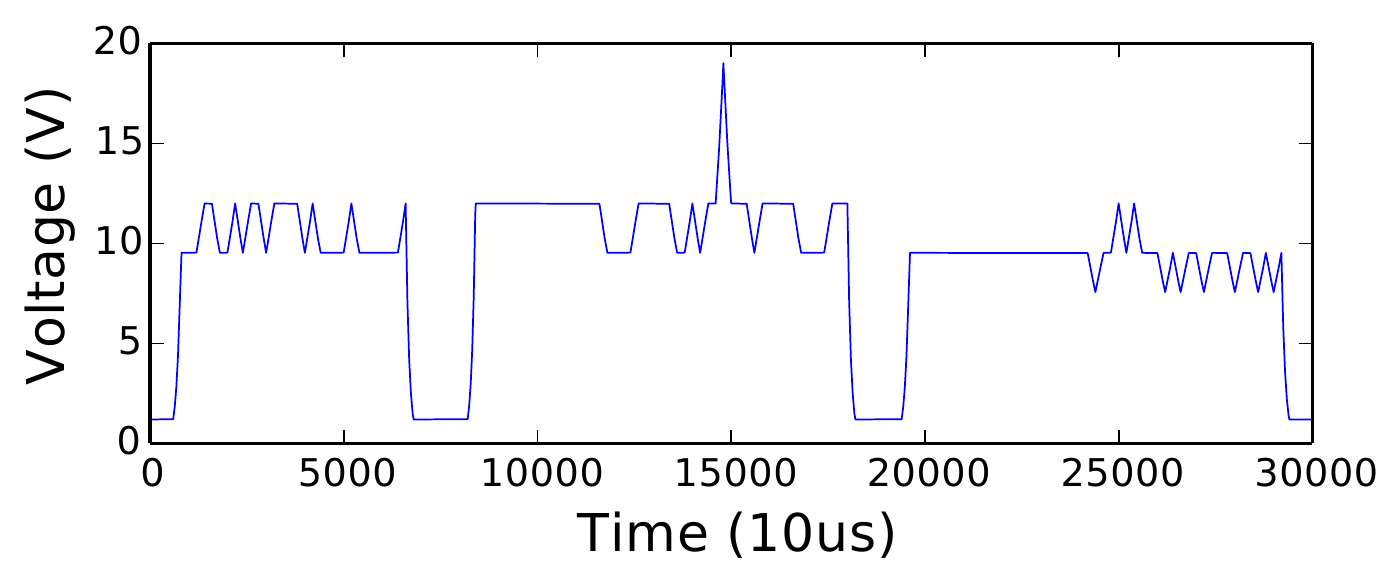}}
\caption{
Energy harvesting traces showing voltage input fluctuations in two different places within about 250$\sim$400ms from an RF energy harvesting reader~\cite{gu2016nvpsim}.
}
\label{fig:trace}
\end{figure}

We use 8 applications in Mibench~\cite{guthaus2001mibench} and
15 applications in Mediabench~\cite{lee1997mediabench} benchmark suites~\cite{liu2017benchprime}.
All the applications are compiled by \name compiler
with -O3 optimization level.
To evaluate \name for realistic energy harvesting environment with frequent power outages, we use
two power traces of the NVPsim which were collected from real RF
energy harvesting systems~\cite{gu2016nvpsim}. 
Figure~\ref{fig:trace} describes
the shape of those two power traces; (a) shows the voltage fluctuations across
time in home, and (b) shows those in office. Trace 2 (office) has more power outages than Trace 1 (home);
\revision{in every 30 seconds, Trace 1 and 2 incur $\approx$20 and $\approx$400 power outages, respectively.}
%

\subsection{Performance Comparison}\label{sec:eval:perf}


\begin{figure*}[htb!]
  \includegraphics[width=1.0\textwidth]{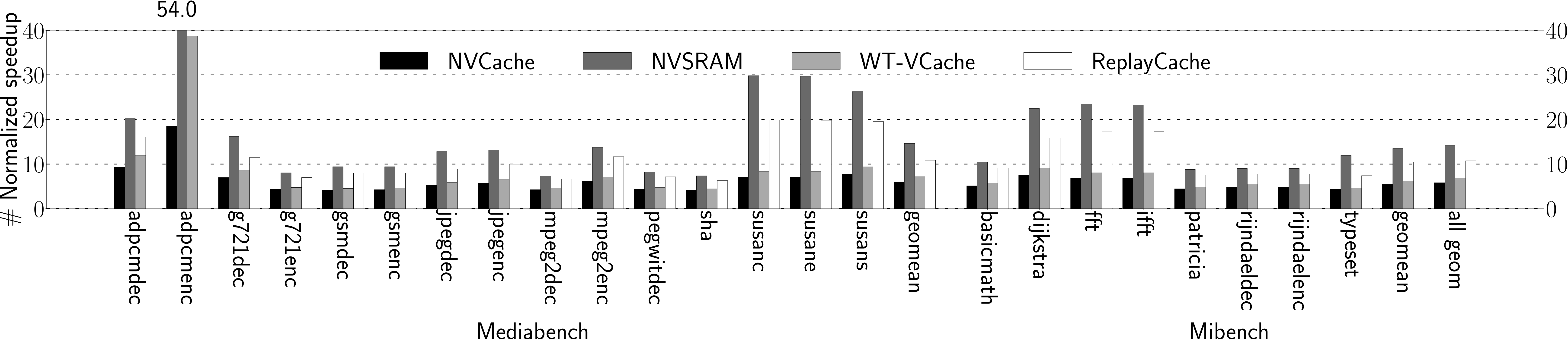}
  \vspace{-15px}
  \caption{Performance results ``without'' power outages. 
  We compare \name with NVCache, NVSRAMCache, and WT-VCache.
  Y-axis shows the normalized speedup over the baseline without a cache.
  The higher, the faster. 
  \vspace{-10px}
  }
  \label{fig:perf}
\end{figure*}
\begin{figure*}[htb!]
  \includegraphics[width=1.0\textwidth]{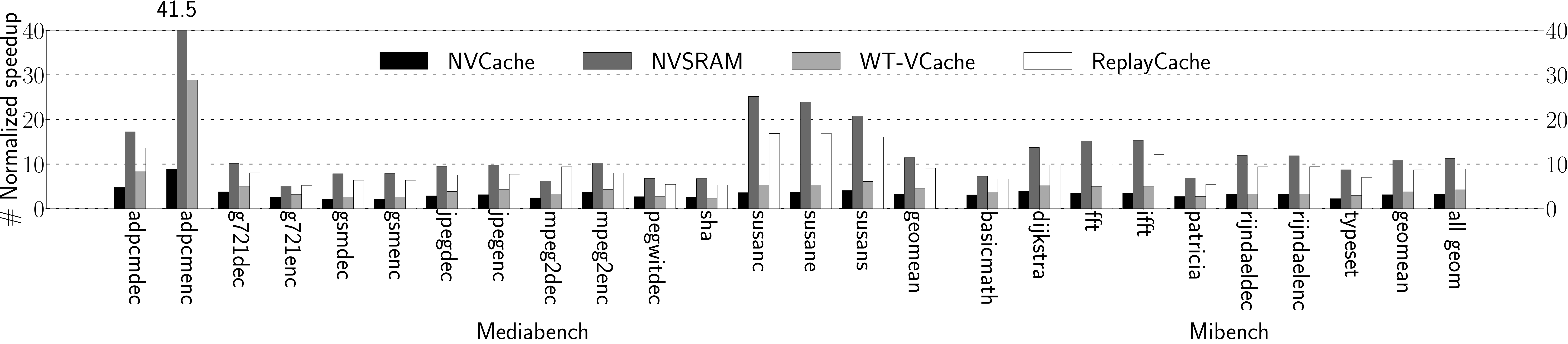}
  \vspace{-15px}
  \caption{Performance results ``with'' power outages, simulated with Power Trace 1 in Figure~\ref{fig:trace}(a). 
  We compare \name with NVCache, NVSRAMCache, and WT-VCache.
  Y-axis shows the normalized speedup over the baseline without a cache.
  }
  \vspace{-10px}
  \label{fig:perf-trace-1}
\end{figure*}
\begin{figure*}[htb!]
  \includegraphics[width=1.0\textwidth]{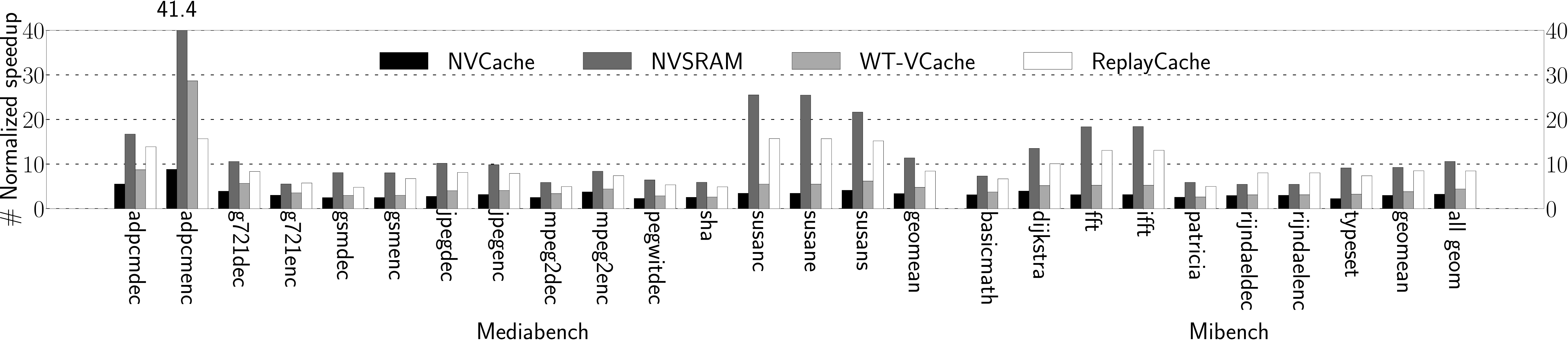}
  \vspace{-15px}
  \caption{Performance results ``with'' power outages, simulated with Trace 2 in Figure~\ref{fig:trace}(b).
  We compare \name with NVCache, NVSRAMCache, and WT-VCache.
  Y-axis shows the normalized speedup over the baseline without a cache.
  }
  \vspace{-10px}
  \label{fig:perf-trace-2}
\end{figure*}



\subsubsection{Performance without Power Outage}
\cref{fig:perf} shows the performance results of power-failure-free executions.
The Y-axis shows the normalized speedup over the baseline without a cache.
\ignore{Although all the schemes here use NVP, NVP+\name and QuickRecall+\name
produce the same results as their register checkpointing are never triggered without power failure.}
Overall, \name improves the performance of all the applications, achieving 11x speedup on (geometric) average. 
It turns out that NVCache is the worst design as expected because of higher latency (especially stores)
then SRAM, but it still improve the performance due to locality exploitation.

Recall that NVSRAM uses a traditional SRAM cache with an NVM (advanced ReRAM) backup, and checkpoints/restores the whole cache state to/from the NVM backup across power failure. Thus, with no power outage, NVSRAM should perform as an original write-back volatile cache. NVSRAM performs the best as expected achieving 14x speedup compared to the baseline. Here, the performance gap between NVSRAM and \name 
results from the store write-back latency that our region-level persistence did not manage to fully hide with ILP.
Later in~\cref{sec:eval:ilp}, we present the detailed results on \name's ILP efficiency, reflecting the amount of stalls at the region boundary.

WT-VCache shows some improvement over the baseline without a cache.    
The performance benefits mostly come from load hits, though the write-through policy makes the cost of store the same as the baseline. \name outperforms WT-VCache, \ie achieving an average speedup of 1.57x, by hiding the latency of stores with region-level persistence.

\subsubsection{Performance with Power Outages}

\cref{fig:perf-trace-1,fig:perf-trace-2} show the performance results with
power failures, simulated on Power Traces 1 and 2 in Figure~\ref{fig:trace}.
The simulation includes different sequences of power up/down and downtime
during charging. Again, the Y-axis is the normalized speedup over the baseline
without a cache.

Although NVCache uses the same NVM technology as main memory, it can be placed close to a core as cache in that core-to-NVCache access is faster than core-to-NVM one.  NVCache remains the worst mainly due to a long cache access latency and higher energy consumption of NVM access wasting hard-won energy.

With power outages, \name achieves $\approx$80\% performance of NVSRAM. This is a promising result given that \name is a software-only scheme that allows commodity systems to use a volatile data cache as is with no other additional hardware support.
Note that 
NVSRAM cache can retain the cache data across a power outage while \name cannot since it uses a traditional SRAM cache that loses all the content upon the outage; due to this advantage, NVSRAM beats all other cache schemes. 
In contrast, when power comes back, \name has to start with a cold cache reloading all necessary data from NVM. 
Nevertheless, the cache warming-up cost can be amortized by the benefit of cache hits, unless the program execution is too frequently interrupted by power failure.

WT-VCache shows only comparable performance to the expensive NVCache design due to the cost of warming up the volatile cache across power failure and serializing stores with the write through policy. However, WT-VCache still outperforms the baseline with exploiting certain degree of locality.
In particular, WT-VCache outperforms \name for {\it adpcmencode}. That is because the \name ended up increasing the instruction count due to a register spilling in a hot loop along with the stack memory access cost.
On average, WT-VCache performance happens to be almost same as NVCache design.

Overall, \name achieves 8.95x (Trace 1) and 8.46x (Trace 2) average speedups compared to the baseline (no cache), outperforming NVCache and WT-VCache.
The reason for the performance gain over them is two-fold. First, \name costs less cache power consumption compared to the NVCache and WT-VCache as shown in \cref{fig:breakdown}.
Second, due to the ILP nature, \name can hide the most of write-back latency as will be shown Figure~\ref{fig:ilp_eff}.

\ignore{
\subsubsection{Cache Miss Rate with Power Outage}
Due to the volatile nature of SRAM cache, all data in SRAM cache will lose when power failure happens for WT-VCache and \name designs. \cref{fig:cache_miss_rate}
shows number of cache miss per 1K instructions across four cache designs with trace 2 used. It clearly indicate that WT-VCache incurs higher miss rate due to more
frequently power failure happened, e.g., WT-VCache suffers 26 miss/1K while other schemes have 14, 17, and 20 miss/1K for VCache, NVSRAMCache, and \name respectively.
}

\vspace{-5px}
\subsubsection{Energy Consumption Breakdown}
\begin{figure}[t!]
  \includegraphics[width=1.0\columnwidth]{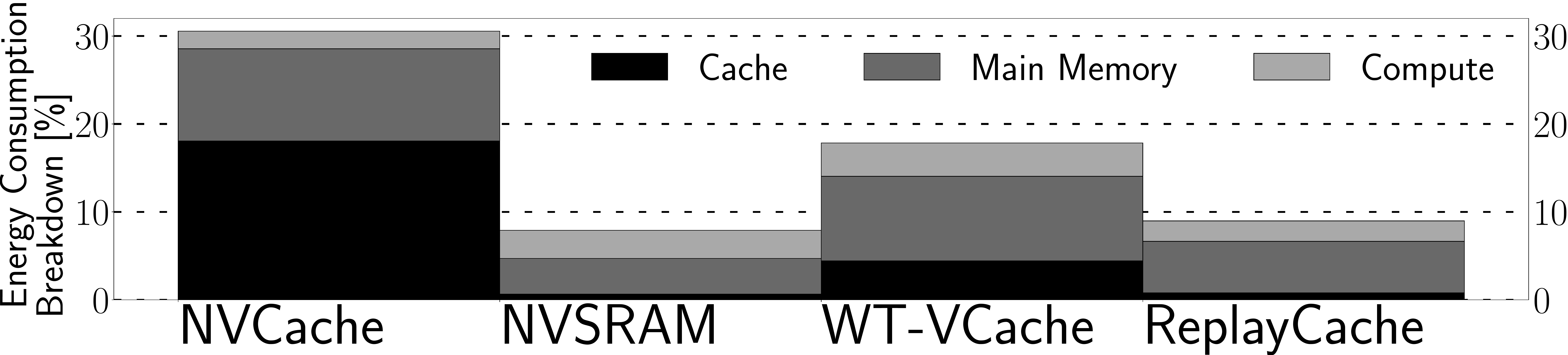}
  \vspace{-15px}
  \caption{\choiadd{Normalized energy consumption breakdown (trace 2) compared to the baseline without a cache.}
  }
  \label{fig:breakdown}
\end{figure}

To figure out the energy consumption behavior of \name, we measured how much energy was consumed
for each part of the system, \ie cache, memory, and core (NVP computation), by using the power
model provided by NVPsim~\cite{gu2016nvpsim}. 
Figure~\ref{fig:breakdown} shows the resulting energy consumption breakdown, normalized to the same no-cache baseline, using the Power Trace 2.
Overall, \name turns out to be very effective, allowing NVP to spend more energy for computation rather than memory access
compared to other schemes. Also, \name's energy consumption is on par with the ideal NVSRAM.
As a result, \name enables NVP to make a significantly further forward progress than the no-cache baseline.

\vspace{-5px}
\subsection{Instruction Level Parallelism Efficiency}\label{sec:eval:ilp}

\begin{figure}[t!]
  \includegraphics[width=1.0\columnwidth]{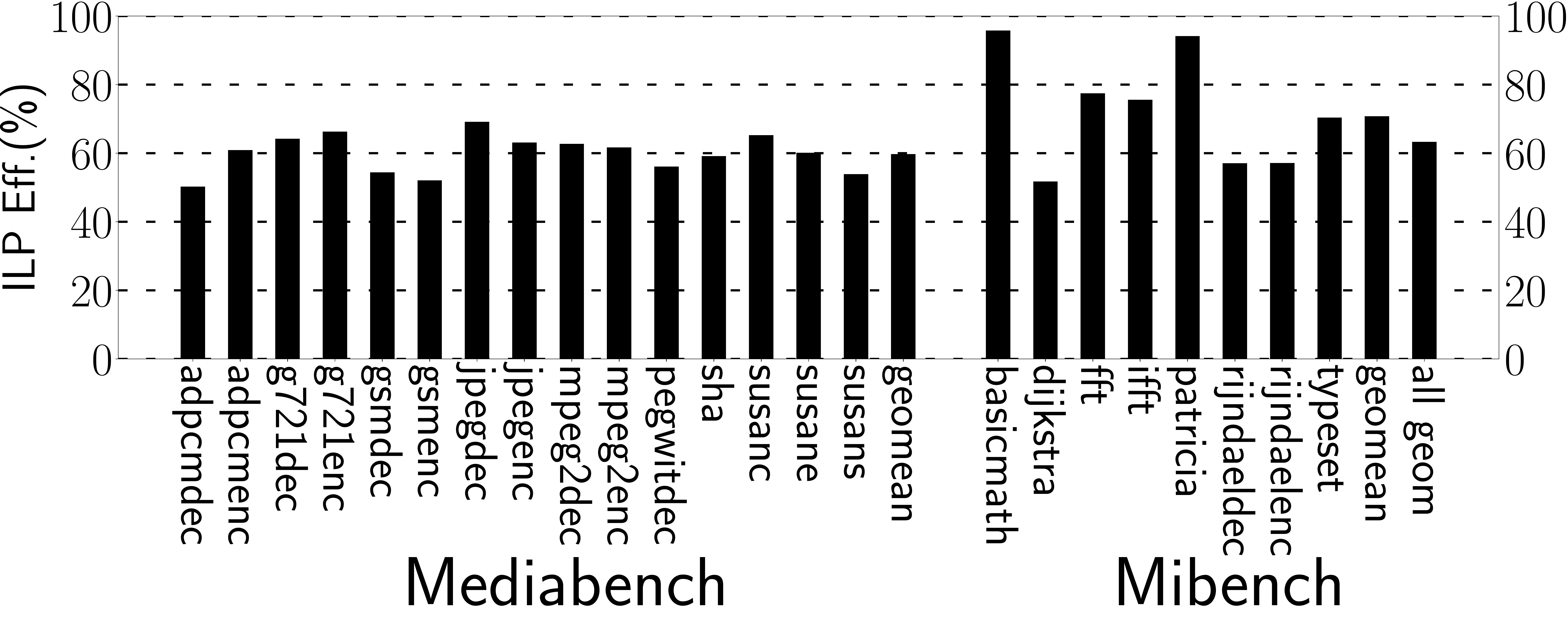}
  \vspace{-15px}
  \caption{Instruction-level parallelism efficiency "without" power failure.}
  \label{fig:ilp_eff}
\end{figure}

\name exploits ILP for stores and thus is faster than a volatile write-through cache.
Nevertheless, its ILP can be bounded by region-level persistence guarantee, \eg a
region end is reached before the preceding store completes the NVM persistence, in
which case \name is slower than an ideal write-back cache. With that in mind, we
investigate the amount of ILP that \name can exploit, based on the power-failure-free
simulation results, to reason about \name's high performance.

Let $N$ be the total number (dynamic instances) of stores in a region.
Among them, $N_{no\_stall}$ represents the number of stores that do not stall, and
$N_{stall}$ represents the number of stores that stall at the region boundary for
region-level persistence guarantee. Let $C$ be the cycles required for a store to
be persisted in the NVM (\ie the write-through NVM store latency; 31 cycles in our
evaluation for default ReRAM); and $S(i)$ be the stall cycles of $i$'s store in the
region. We then calculate the ILP efficiency at a 0-to-100\% scale. For each store,
the worst efficiency 0\% is made when the processor waits for $C$ cycles after the
region finishes, and the best efficiency 100\% reflects 0 stall cycle. \cref{equ:ilp}
defines the ILP efficiency for $N$ stores in a region as follows. 
\vspace{-0.2cm}
\begin{equation}\label{equ:ilp}
ILP_{eff} (\%) = \frac{1}{N}\{\sum_{i=1}^{N_{no\_stall}}1 + \sum_{i=1}^{N_{stall}}(1-\frac{S(i)}{C})\}*100
\end{equation}

\cref{fig:ilp_eff} shows the ILP efficiency of the tested applications. On average,
\name achieves 63\% ILP across the evaluated applications, and the ILP efficiency explains why
\name achieves the performance shown in Figure~\ref{fig:perf}. Again, in our evaluation, the
write-through store latency takes 31 cycles~\cite{gu2016nvpsim}, i.e., $C=31$.  This
implies that \name can hide about 20 cycles out of the 31 cycles on average.

\subsection{Binary Size Analysis}\label{sec:recovery_code_size}
\begin{figure}[htb!]
  \includegraphics[width=1.0\columnwidth]{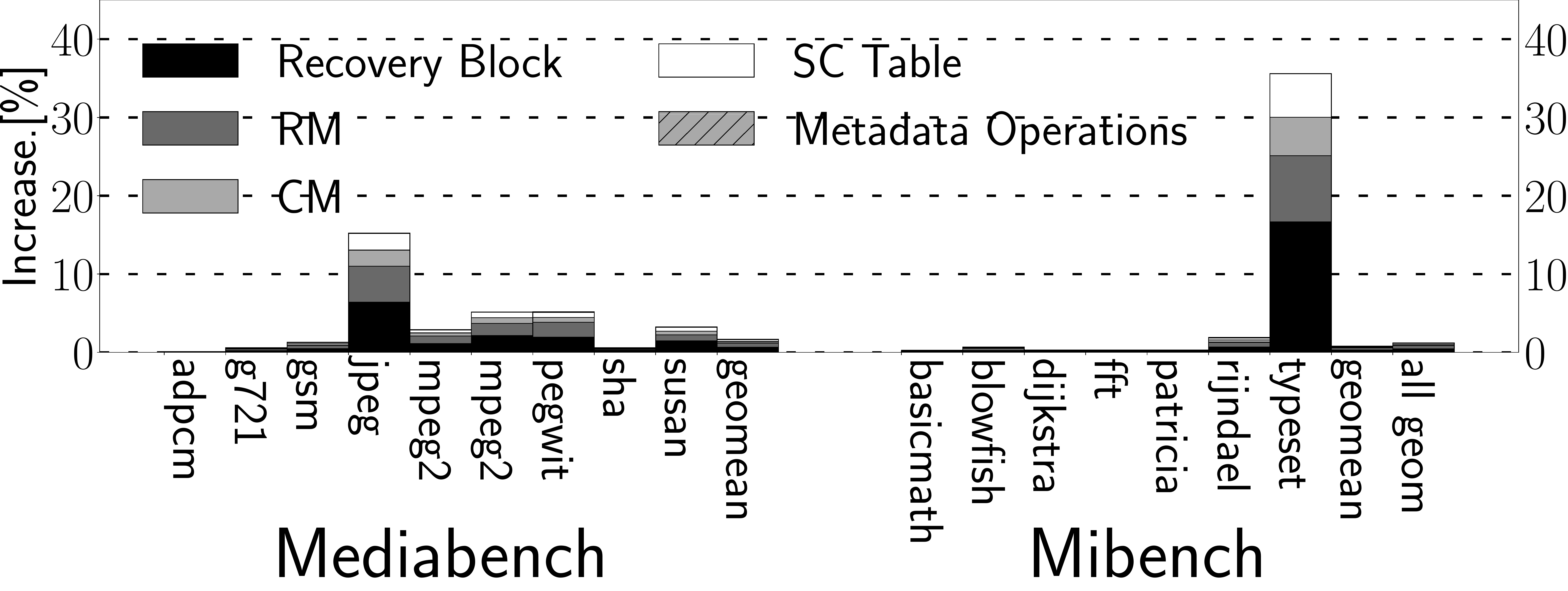}
  \vspace{-15px}
  \caption{\revision{Binary size increase due to recovery block, metadata (RM, CM, SC table), and metadata operations (code).}}
  \vspace{-10px}
  \label{fig:recovery_code_size}
\end{figure}

\revision{
\cref{fig:recovery_code_size} demonstrates the breakdown of binary size increase of \name
binaries as a percentage increase compared to the baseline binary. Overall,
\name incurs only 1.2\% binary size overhead on average. Metadata operations are comprised of roughly 110 instructions, leading to near-zero overhead.
Only 2 applications,
\eg \texttt{jpeg} and \texttt{typeset}, have observable binary size increase because they have lots of small regions. 
Note that 
the binary size overhead never puts pressure on application's memory usage at run time. That is because the metadata is accessed only at boot time on which \name's recovery  starts with empty cache---already wiped out upon the prior failure---without cache pollution.
} 

\revision{
\subsection{Dynamic Instruction Count Analysis}\label{sec:dynamic_inst_count}
}
\begin{figure}[htb!]
  \includegraphics[width=1.0\columnwidth]{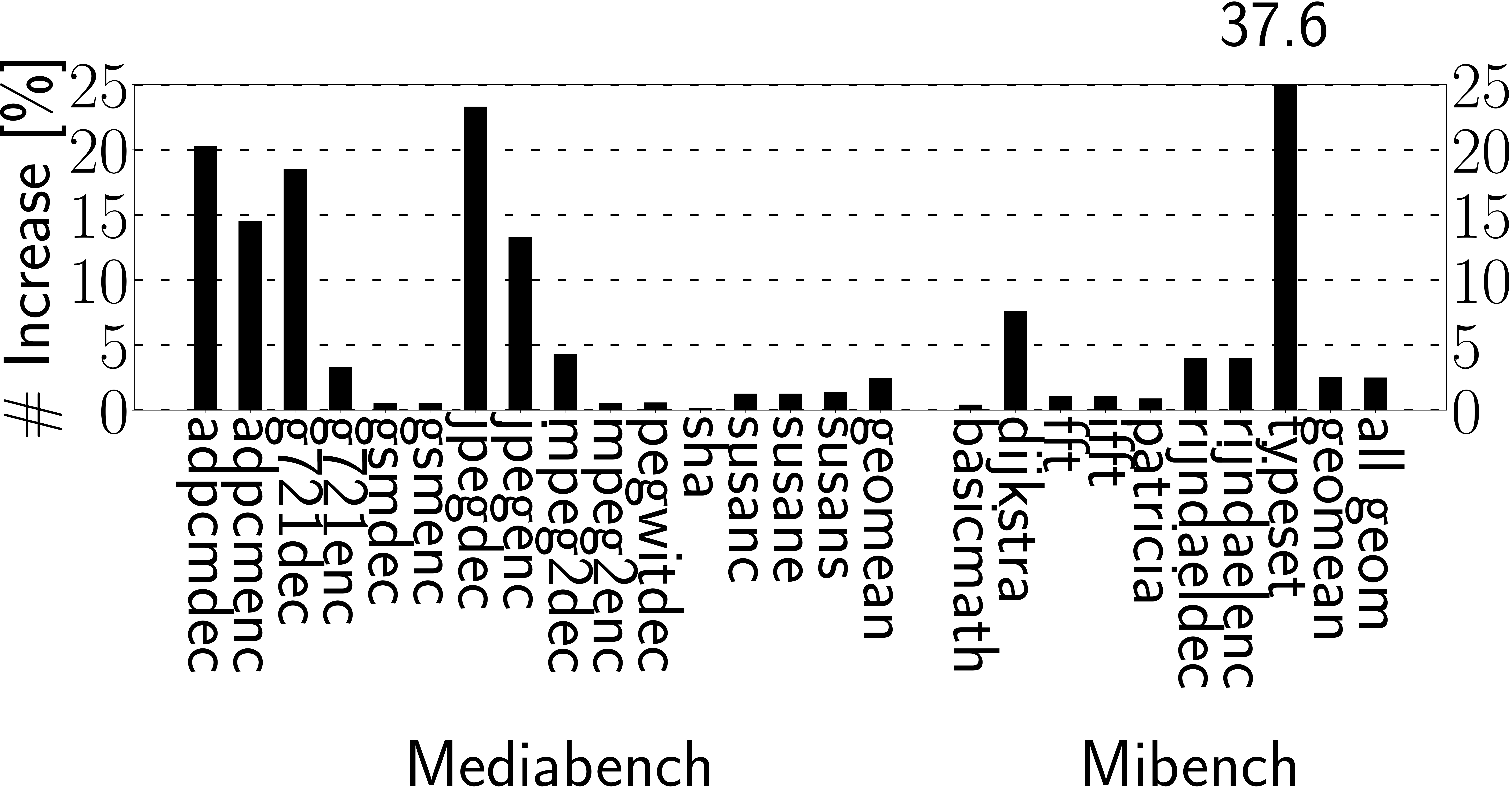}
  \vspace{-15px}
  \caption{\revision{Dynamic instruction count increase due to \name compiler code generation;
  lower is better.}}
  \vspace{-10px}
  \label{fig:dynamic_inst_count}
\end{figure}

\revision{
\cref{fig:dynamic_inst_count} demonstrates that \name compiler only increases dynamic instruction
count by 2.49\% on average compared to the baseline binary. Note that this is not a critical performance limiting factor as confirmed in Figure \ref{fig:perf}-\ref{fig:perf-trace-2} where \name  consistently shows significant speedups.
}

\subsection{Sensitivity Study}\label{sec:eval:sense}


\subsubsection{Cache Size}

\begin{figure}[t!]
  \includegraphics[width=1.0\columnwidth]{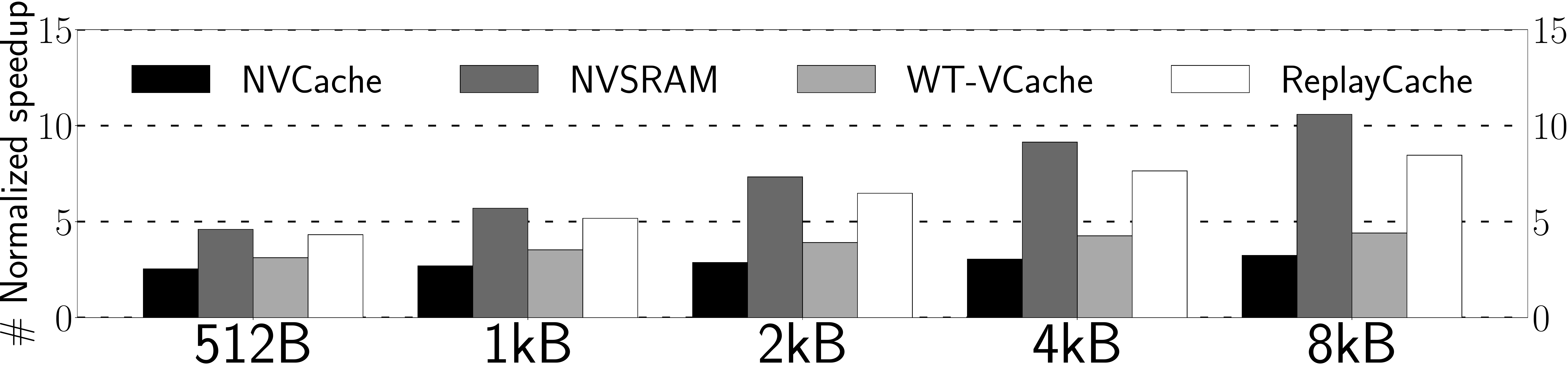}
  \vspace{-15px}
  \caption{Cache size sensitivity analysis for Trace 2.
  \vspace{-5px}
  }
  \label{fig:cachesense}
\end{figure}

\cref{fig:cachesense} shows the normalized execution time (to the baseline without a cache)
of alternative cache schemes with a different cache size from 512B to 8KB using Power Trace 2.
The results show that \name matches the performance of NVSRAM cache (that is an ideal write-back
cache in power-failure-free scenarios) for small cache size, such as 512B and 1KB.

\subsubsection{NVM Technology}

\begin{figure}[t!]
  \includegraphics[width=1.0\columnwidth]{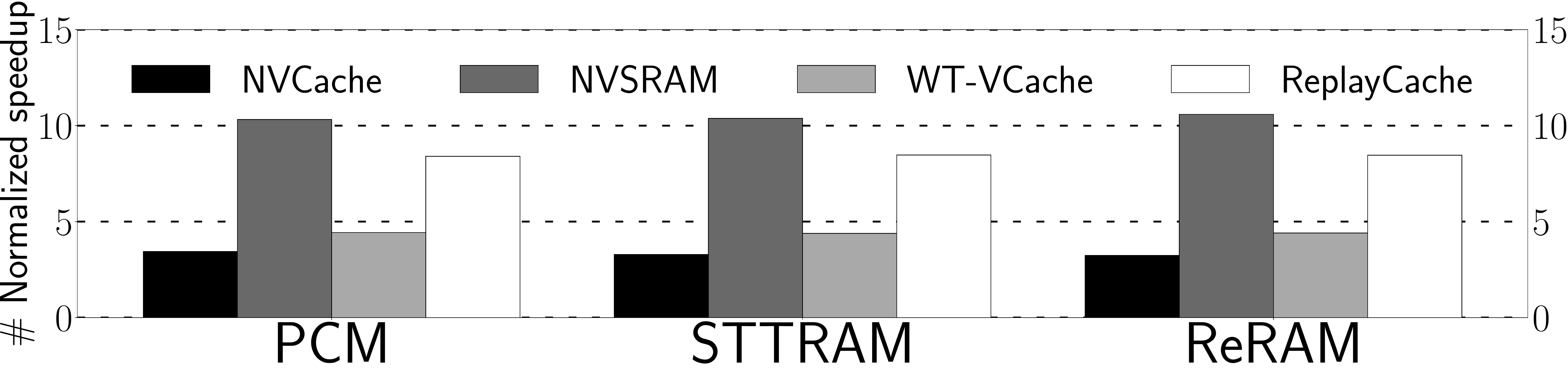}
  \vspace{-15px}
  \caption{Sensitivity study on different NVMs with trace 2 used.
  }
  \vspace{-5px}
  \label{fig:nvm_sense}
\end{figure}

Different NVM technologies (\eg, ReRAM, PCM, and STT-RAM) have different write/read
latency properties as summarized in~\cref{table:NVM}.
For ReRAM, PCM, and STT-RAM (as the main memory), Figure~\ref{fig:nvm_sense} shows the
normalized speedup of alternative cache schemes, compared to their 3 baselines without a cache. 
It turns out that \name consistently achieves significant speedups across the NVM technologies (8.4x-8.46x).

\subsubsection{NVP versus QuickRecall}
\begin{figure}[t!]
  \includegraphics[width=1.0\columnwidth]{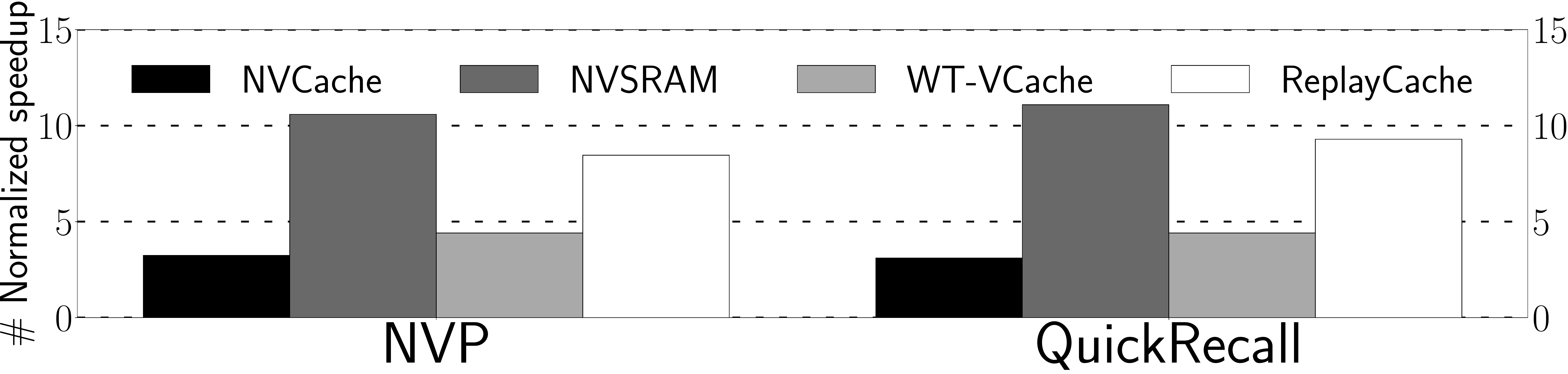}
  \vspace{-15px}
  \caption{Performance overhead comparison with trace 2.
  }
  \label{fig:quickrecall-trace-2}
\end{figure}
To analyze the impact of the underlying just-in-time register checkpointing on \name's
performance, we tested all four cache schemes on top of QuickRecall and compared the results with those of NVP.
Again, we used the Power Trace 2 and normalized the speedup over their baselines, \ie NVP/QuickRecall without cache. 
Figure~\ref{fig:quickrecall-trace-2} describes that the performance trend is similar to NVP;
however, it is worth noting that QuickRecall requires higher checkpoint/restoration voltage
due to data backup as shown in \cref{table:comp}---though it is a less expensive system than NVP due to the lack of non-volatile flip-flops.

\ignore{
As this paper discussed, QuickRecall checkpoints registers to NVM at power-off time without having the NVFF. 
Since this checkpointing delay becomes longer, QuickRecall must increase the checkpoint voltage level or having a large size energy reservoir to achieve failure-atomic register checkpointing.
For QuickRecall, we increased the checkpoint voltage level as shown in Table~\ref{table:comp}.
}

\subsection{\name Compiler Region Statistics}\label{sec:eval:region}

\begin{figure}[t!]
  \includegraphics[width=1.0\columnwidth]{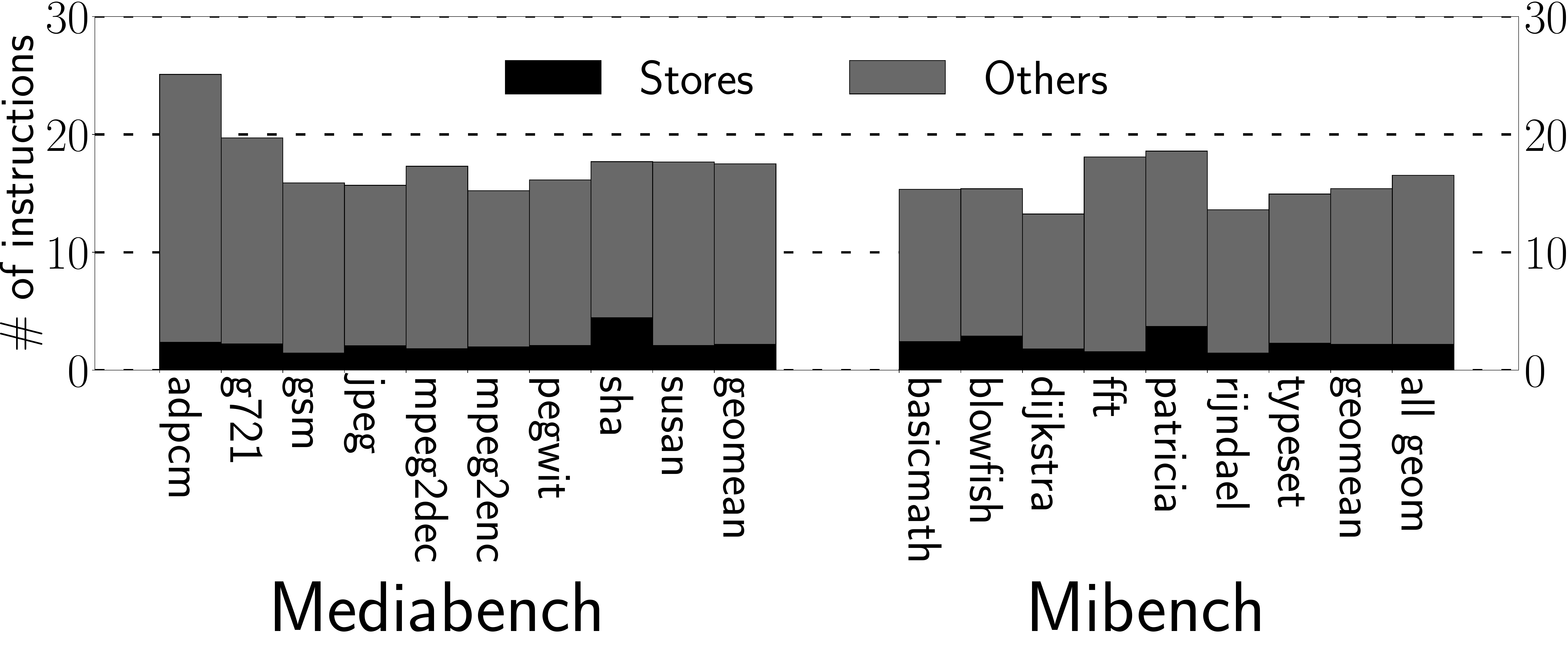}
  \vspace{-15px}
  \caption{Breakdown of per-region instructions on average.
  }
  \label{fig:region_char}
  \vspace{-5px}
\end{figure}

\begin{figure}[t!]
  \includegraphics[width=1.0\columnwidth]{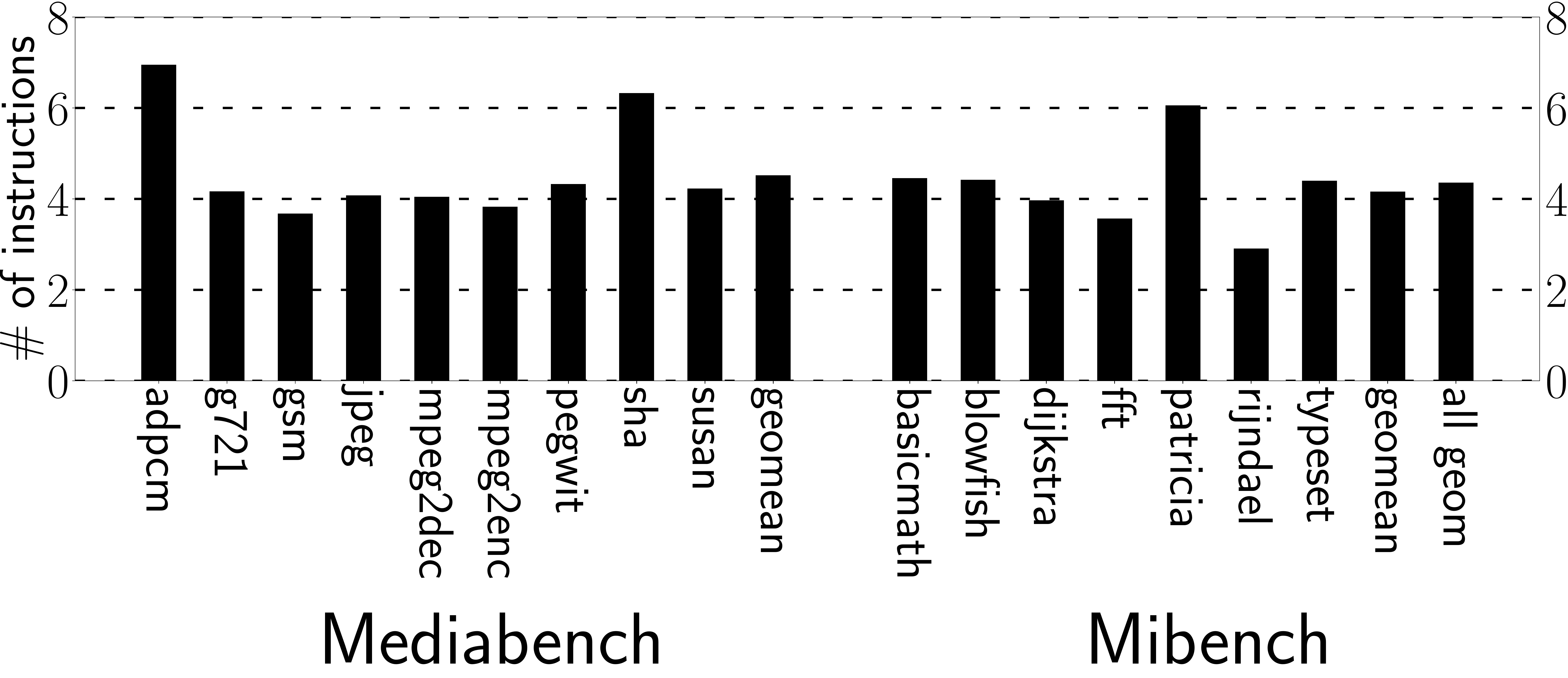}
  \vspace{-15px}
  \caption{Average distance (the number of instructions) between region's last store and the following region boundary.
  }
  \vspace{-5px}
  \label{fig:store_distance}
\end{figure}

We study the region statistics, statically calculated from the binary built by our compiler.
Figure \ref{fig:region_char} presents the average number of instructions per region. 
On average, there are 16.4 instructions per region.  We also break them down into two categories: stores and other instructions.  On average, there are 2.18 stores and 14.35 others per region.
This implies that the recovery code blocks are not long either (smaller than their regions). In fact, we did not encounter any recovery block that requires the corresponding region to be split to ensure the absence of power failure during the recovery.

Moreover, Figure \ref{fig:store_distance} shows the average distance (the number of instructions)
between the last store of a region and the following region boundary, i.e., 4.35 instructions on average. The distance here reflects \name's ILP 
opportunities.

%% file: related-works.tex
\section{Related Works}

Many prior works~\cite{mittal2014lastingnvcache,agarwal2019improving,jokar2015sequoia,park2012future,xu2009design,wang20132,mittal2014writesmoothing} have been
proposed to leverage non-volatile caches to speed up the performance and
leverage their zero standby leakage and crash consistency free properties.
However, the cell endurance of NVM techniques ranges from $10^5$ in flash to
$10^{12}$ in STT-RAM. Non-volatile caches may only be able to endure
few months for most of real applications~\cite{jokar2015sequoia}. Thus, prior
works focus on increasing the lifetime of NVM cells. Furthermore, NVM
has the asymmetric performance property. A write is considerably
slower than a read, compared to the SRAM counterpart. Both
the short lifetime and the long write latency severely limit the use of NVM as
L1 cache in practice.

To use the synergy of NVM and SRAM, many
researches~\cite{singh2019design,majumdar2016hybrid,herdt1992analysis,liu2020low,chiu2012low,miwa2001nv,masui2003design,sheu2013reram,lee2015rram,yamamoto2009nonvolatile,li2017design,singh2019design} proposed to
incorporate different NVM technologies (\eg STT-RAM, ReRAM, etc.) with SRAM.
Many proposals leverage the NVM part as a just-in-time checkpointing storage of the traditional SRAM-based cache in case of power failure. Thus, the NVM speed is the critical aspect for the success of such SRAM/NVM hybrid design.
Although researchers attempt to improve the NVM backup/restoration latency~\cite{li2017design,singh2019design}, they assume forward-looking technologies; no current NVM technologies provide comparable latency to SRAM~\cite{kolli2017architecting,choi2019cospec}. 

The idea of partitioning a program into multiple regions to design more
efficient energy harvesting systems has been explored.
Ratchet~\cite{van2016intermittent} proposed to partition program into a series
of anti-dependence-free (\ie write-after-read dependence free) regions for idempotent processing as with others~\cite{de2012static,de2013idempotent,liu2018ido,liu2016lightweight,liu2015clover,liu2016compiler,liu2017compiler,kim2020compiler}. 
Since idempotent regions can be safely
re-executed multiple times, it can recover a power-interrupted region by rolling back to the beginning in the wake of power failure, provided the inputs value of the region can survive the power failure. Due to the absence of the anti-dependence, Ratchet only needs to checkpoint all live-in registers of the region at its entry point. Unfortunately, such consecutive NVM writes are not only expensive but also dangerous increasing the chance of power failure in the middle of their writes.
To address the issues in Ratchet, Clank~\cite{hicks2017clank} proposed hardware-based
idempotent processing. Despite its improved performance, Clank requires
relatively heavy and complex hardware components such as a fast scratchpad
memory for speeding up the writes to the underlying NVM and an expensive CAM
(content-addressed matching) search based load/store address tables to dynamically detect anti-dependence.
Alternatively, CoSpec~\cite{choi2019cospec} proposed power failure speculation assuming that power failure is not likely to occur. Thus, it buffers all the application writes in a gated store buffer~\cite{liu2016low,zeng2021turnpike} in case of misspeculation, \ie actual power failure. Also, the CoSpec compiler partitions program into a series of regions so that they never overflow the store buffer. When power failure occurs in the middle of a region, it is rolled back to the beginning in the wake of power failure. As with Ratchet, CoSpec needs to pay the overhead of checkpointing all live-in registers of every region.
Unlike \name, neither Clank nor CoSpec supports a volatile data cache. Thus, we suspect that \name can significantly outperform them.

%% file: conclusion.tex
\section{Conclusion}
This paper presents \name, a software-only scheme that
enables energy harvesting systems to take advantage of
a volatile data cache efficiently and correctly. To
achieve crash consistency with the volatile data cache,
\name proposes a replay-based solution that 
restores the operands of potentially unpersisted 
stores from the register checkpoint and then re-executes
them to restore consistent non-volatile memory status.
Experimental results show that compared to the baseline
with no cache, \name significantly improves the performance
by 8.46x-\revision{8.95x} speedup on geometric mean, while
ensuring correct resumptions even in the presence of 
unpredictable and frequent power outages.